\documentclass[10pt]{IEEEtran}
\usepackage{ifpdf}
\ifCLASSOPTIONcompsoc
\else
\fi
\ifCLASSINFOpdf
   \usepackage[pdftex]{graphicx}
   \DeclareGraphicsExtensions{.pdf,.eps}
\else
  \usepackage[dvips]{graphicx}
  \DeclareGraphicsExtensions{.eps}
\fi
\ifCLASSOPTIONcompsoc
  \usepackage[caption=false,font=normalsize,labelfont=sf,textfont=sf]{subfig}
\else
  \usepackage[caption=false,font=footnotesize]{subfig}
\fi
\begin{document}
\title{Evolution of the Internet $k$-dense structure}

\author{Chiara~Orsini%
\thanks{C. Orsini is the corresponding author. She is with CAIDA/UCSD, IIT/CNR and IET/UNIPI. e-mail: chiara@caida.org },%
~Enrico~Gregori\thanks{E. Gregori is with the Institute of Informatics and Telematics, Italian National Research Council (IIT/CNR), Pisa, Italy. e-mail: enrico.gregori@iit.cnr.it},%
~Luciano~Lenzini\thanks{L. Lenzini is with the Department of Information Engineering, University of Pisa (IET/UNIPI), Pisa, Italy. e-mail: l.lenzini@iet.unipi.it},%
~Dmitri~Krioukov\thanks{D. Krioukov is with the Cooperative Association for Internet Data Analysis, University of California San Diego (CAIDA/UCSD), San Diego, CA, USA. e-mail: dima@caida.org}}% <-this % stops a space

\IEEEtitleabstractindextext{%
\begin{abstract}
%\boldmath
As the Internet AS-level topology grows over time, some of its structural properties remain unchanged. Such time-invariant properties are generally interesting, because they tend to reflect some fundamental processes or constraints behind Internet growth. As has been shown before, the time-invariant structural properties of the Internet include some most basic ones, such as the degree distribution or clustering. Here we add to this time-invariant list a non-trivial property---$k$-dense decomposition. This property is derived from a recursive form of edge multiplicity, defined as the number of triangles that share a given edge. We show that after proper normalization, the $k$-dense decomposition of the Internet has remained stable over the last decade, even though the Internet size has approximately doubled, and so has the $k$-density of its $k$-densest core. This core consists mostly of content providers peering at Internet eXchange Points, and it only loosely overlaps with the high-degree or high-rank AS core, consisting mostly of tier-1 transit providers. We thus show that high degrees and high $k$-densities reflect two different Internet-specific properties of ASes (transit versus content providers). As a consequence, even though degrees and $k$-densities of nodes are correlated, the relative fluctuations are strong, and related to that, random graphs with the same degree distribution or even degree correlations as in the Internet, do not reproduce its $k$-dense decomposition. Therefore an interesting open question is what Internet topology models or generators can fully explain or at least reproduce the $k$-dense properties of the Internet.
\end{abstract}

\begin{IEEEkeywords}
Internet topology, network evolution, $k$-dense, $dK$-graphs.
\end{IEEEkeywords}}

\maketitle

% is not selected <OR> if conference mode is selected - because compsoc
% conference papers position the abstract like regular (non-compsoc)
\IEEEdisplaynontitleabstractindextext

% \fi
\IEEEpeerreviewmaketitle

\section{Introduction}
\label{sec:introduction}

The discovery of power laws in the Internet in 1999 \cite{FALOUTSOS} came as a big surprise to many, and as a source of major disbelief to some. Even more surprising is that over the last decade, an increasing number of increasingly refined and complete macroscopic Internet topology measurements and data sources (with one exception, WHOIS) show that the power-law distribution of AS degrees in the Internet has remained exceptionally stable, i.e.\ time-invariant \cite{PLAWTON, TWELVE, PSBOOK, ITT,TOPOCOMP, LIXIAPOW}.

One has to always keep in mind that these macroscopic measurements may miss a significant percentage of links. Indeed, there have been several studies supporting this expectation: huge percentages of links are reported as missing in \cite{ASREL,LOL,ELUSIVEGT,LARGEIXP,FEEDER}, for example. These studies rely on proprietary data collected from only a few ASes. Therefore questions about statistical significance of the reported results, which are difficult or impossible to reproduce due to the proprietary nature of the data, are well grounded. However these questions may be not so important in view of that all these studies report that a majority of missing links are peering, while customer-provider links appear to be covered well in macroscopic topology measurements. These results are expected, given an ease at which peering links are ``set up'' at Internet eXchange Points (IXPs), for example: as soon as an AS connects to an IXP and declares an open peering policy, it can exchange traffic with any other openly peering AS at the same IXP. It is instructive to compare the setup of such ``links'' with the process behind setting up customer-provider links, often involving complicated business decisions, negotiations, and payment agreements \cite{PEERINGECONOMIES,NORTONPEERING}. In other words, peering links are definitely more volatile than customer-provider links. At the same time, due to the specifics of Internet topology measurements, the higher in the customer-provider AS hierarchy are the two ASes connected by a peering link, the higher the probability that topology measurements detect this link. Therefore a majority of missing peering links are between peripheral ASes, and they might affect the results presented here as discussed in Section~\ref{sec:incompleteness}.

The time-invariant nature of the power-law distribution of (customer-provider) degrees, as well as of strong clustering, has recently found a new explanation as a consequence of  trade-off optimization between AS popularity and similarity in Internet evolution \cite{POPSIM}. This optimization is rather general and applies not only to the Internet, but also to social and biological networks, and even to spacetime in our accelerating universe \cite{UNIVERSE}. More generally, any time-invariant property of an evolving network structure is a candidate to reflect some fundamental forces or constraints behind network evolution, unless this property is a simple statistical consequence of some other property---degree distribution, for example. These forces and constraints can be quite general, applicable to many different networks, as is the case with degree distribution and clustering, or they can be specific to a given network, in which case they likely reflect some specific functions that this network performs.

\begin{figure}[ht]
\centering
\includegraphics[width=0.4\textwidth]{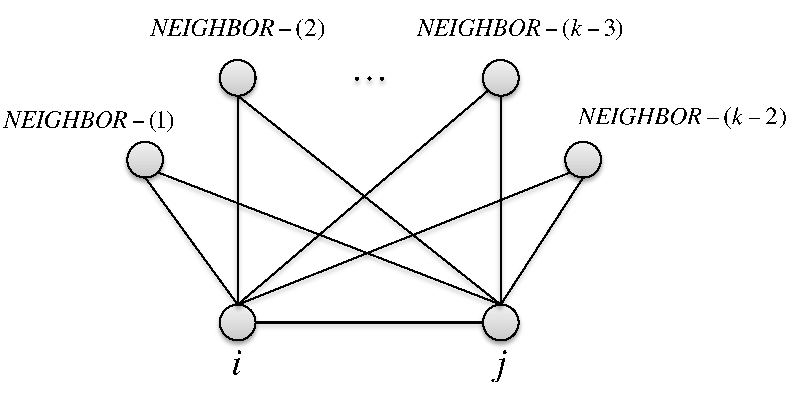}
\caption{Link $i,j$ of multiplicity $m(i,j)=k-2$. Edge multiplicity is the number of neighbors common to the two nodes that this edge connects \cite{OURDENSE}.}
\label{fig:kdef}
\end{figure}

Here we show that the Internet $k$-dense decomposition is time-invariant, and explain this invariance via Internet-specific AS data reflecting different functions and business roles of different ASes.  The $k$-dense decomposition~\cite{KDENSE} of a graph $G$ is a hierarchy of nested subgraphs $H_k$, $H_k\subset H_{k-1} \subseteq G$, $k=3,4,\ldots,k_{MAX}$, induced by edges belonging to $k-2$ or more triangles within $H_k$. When $k=2$, $H_k$ is equal to $G$. A more practical definition of $k$-dense is the following: a link $(i,j)$ belongs to a $k$-dense subgraph $H_k$ if and only if nodes $i$ and $j$ that this link connects have at least $k-2$ common neighbors within $H_k$, Figure \ref{fig:kdef}. If a link belongs to $H_k$ but not to $H_{k+1}$, it is said to have the $k$-dense-index equal to $k$. The $k$-dense-index of a node is the maximum $k$-dense-index of its incident edges.
%All further definitions are in Section~\ref{sec:methods}.

The $k$-dense-index of an edge is thus a recursive variant of edge multiplicity, defined as the number of triangles in $G$ that an edge is a member of. Edge multiplicity was introduced and studied in \cite{EMULTI,NEMULTI}, where it was also shown that the edge multiplicity distribution in many real networks is either power-law or fat-tailed, and that many existing network models fail to reproduce this property. In  \cite{POLCLUST} the authors study the $k$-dense decompositions of different networks to scrutinize the global organization of clustering in them. The authors find that in most networks, triangles overlap similarly to how they overlap in random graphs with the same node degree and clustering distributions, and not as in more structured network models with modular organization. In~\cite{DENSE_ANALYSIS,OURDENSE} the authors perform a thorough analysis of the $k$-dense decomposition of a single Internet topology snapshot. They find that the $k_{MAX}$-dense ASes play an important role in Internet connectivity ($43\%$ of AS links in the Internet are attached to these ASes), and that there exist some correlations between the presence of dense subgraphs in the Internet AS-level topology and: a)~ASes participating in one or more IXPs and, b)~ASes with a wide geographical scope (i.e.\ ASes that have points of presence in more than one continent). Compared to this work, the high-level summary of our main results here is:
\begin{itemize}
\item we focus on \emph{evolution} of the Internet's $k$-dense decomposition: we base our study on a series of Internet topology snapshots collected from May 2004 to May 2012, and find that some important $k$-dense properties are time-invariant over the entire period;
\item we show that the Internet's $k$-dense properties are statistically significant using the $dK$-series methodology~\cite{DKGRAPHS};
\item we collect and use data related to IXPs, business relationships, and organization of customer-provider hierarchies to understand the drivers behind the formation of AS communities with different $k$-densities;
\item we find some dependencies between peering policies and complexity of the $k$-densest core in the Internet.
\end{itemize}
In more detail, our main results are:
\begin{enumerate}
\item The maximum $k$-dense-index $k_{MAX}$ exhibits a clear growing trend, increasing from $29$ in 2004 to $48$ in 2012, Section~\ref{sec:trends}, while the number of ASes in the densest core $H_{k_{MAX}}$ is small and fluctuating---it is $59$ in 2004 and $60$ in 2012. The $k_{MAX}$ growth can only partially be attributed to the growing average degree. Other factors are likely related to increasing open peering at IXPs: all the $60$ ASes in the 2012 $H_{k_{MAX}}$ are present at at least one IXP, Section~\ref{sec:kmax-growth}.
\item After proper normalization, Section~\ref{sec:normalization}, the Internet $k$-dense decomposition is shown to be time-invariant in Section~\ref{sec:kdd}. The Internet-specific interpretation of this result appears in Sections~\ref{sec:23-dense} and \ref{sec:kmax-dense}.
\item A significant part of this interpretation is centered around the observation that the degree of an AS and its $k$-density reflect two different Internet-specific properties of the AS, two different AS business roles. Related results include:
\begin{enumerate}
\item While some correlations between AS degree and $k$-density are present as expected, the relative fluctuations are strong, i.e.\ nodes with the same degree or $k$-density have high-variance distributions of their $k$-densities and degrees, respectively, Section~\ref{sec:density.vs.degree}.
\item Random graphs having the same degree distribution or even degree correlations as the Internet, do not have the same $k$-dense decomposition, Section~\ref{sec:dk-analysis}.
\item The analysis of PeeringDB data reveals that while high-degree or high-rank ASes tend to be tier-1 transit providers, high-$k$-density ASes tend to be tier-2 and content providers peering at IXPs, Section~\ref{sec:kmax-dense}.
\end{enumerate}
\item The structure of densest core $H_{k_{MAX}}$ is statistically determined by its degree distribution alone, Section~\ref{sec:kd}. We interpret this result via open peering policies of participating ASes, which do not choose their peers based on their degrees, Section~\ref{sec:1k}. On the contrary, selective peering policies, present elsewhere in the Internet, likely introduce degree correlations, providing a new interpretation of the main result in~\cite{DKGRAPHS}, where the AS-level topology of the Internet was found to be statistically determined by its degree correlations.
\end{enumerate}

Some other {\bf related work} dealing with Internet evolution includes \cite{ITT}, which evaluates how different structural properties, related to the distributions of node degrees, centralities, path lengths, community structure, etc., change over time, from January 2002 to January 2010. The study relies on the Cram\'{e}r-von Mises criterion to identify changes between the distributions, and finds that most distributions remain unchanged, except for the average path length and clustering coefficient. These changes are interpreted as a consequence of peering policy changes. The different growth dynamics of the IPv4 and IPv6 topologies from 1997 to 2009 are juxtaposed in \cite{IP4IP6}. The main result is that IPv4 topology growth had a phase transition in 2001, while IPv6 had a different phase transition in 2006. The authors of \cite{ASEVOLUTION} focus on topology liveness and completeness problems, comparing different Internet topology measurement data sources for the period from January 2004 to December 2006. Two evolution trends are highlighted in this work: a)~customer networks are the major cause of the overall topology growth, b)~transit providers tend to form increasingly denser structures. The monumental study \cite{TWELVE} analyzes the evolution of customer-provider connections in the Internet from January 1998 to January 2010. AS links and nodes are labeled by business relationships and roles, and studied separately. The authors find that enterprize networks and content/access providers at the periphery are the main contributors to the overall Internet growth. They also study rewiring activity, and find that content/access providers appear as most active in that regard.
The metrics that are perhaps the most similar to the $k$-dense decomposition are the $k$-core decomposition~\cite{KCORE,JELLY,FINDCORE,VESPIGNANI} and the $k$-clique percolation method for community detection~\cite{KCLIQUE}. A $k$-core of a graph is its subgraph whose node degrees within the subgraph are greater or equal to~$k$, and a $k$-clique community is a maximal union of adjacent $k$-cliques, where two $k$-cliques are adjacent if they share $k-1$ nodes.
The evolution of the $k$-core decomposition of the Internet from December 2001 to December 2006 is studied in~\cite{COREEVOLUTION}. Similar to the $k$-dense decomposition that we analyze here, the $k$-core decomposition also appears time-invariant according to~\cite{COREEVOLUTION}. However, contrary to the maximum $k$-density, the maximum $k$-coreness does not exhibit any growing trend. To the best of our knowledge, it remains unclear if these results in general, and AS coreness in particular, can find any Internet-specific interpretations.

\section{$k$-dense method}
\label{sec:methods}

\subsection{Motivations}

The $k$-core and $k$-dense decompositions, as well as the $k$-clique percolation method~(CPM), all identify densely connected communities that may still have many connections going outside, contrary to the modularity maximization~\cite{GIRVANNEWMAN}, for example, where the number of such outside connections is minimized. We believe that these $k$-methods suit better the AS Internet reality, where for instance a group of large Internet Service Providers~(ISPs) in a common geographical region are likely to be highly interconnected and thus form a dense community, but at the same time the same ISPs may have a large number of connections directed outside of this community to their dispersed customers~\cite{MYPT}.

The three $k$-methods are similar also in that they all are deterministic, recursive, and provide a set of nested communities of increasing density. Yet they are different in how this density is defined. Any node in a $k$-core has at least $k$ connections to other nodes in the same $k$-core; connected nodes in a $k$-dense community share at least $k-2$ common neighbors and hence overlapping triangles belonging to the same community; while CPM communities are groups of overlapping maximal $k$-cliques---see Figure~\ref{fig:kcomparison} for a basic comparison of these decompositions of a small sample graph. A more detailed comparison of these $k$-methods can be found in~\cite{MYPT}.

We believe the $k$-dense decomposition is optimal among the three methods. Due to its common-neighbor requirement, the $k$-dense approach suggests a stronger relationships among nodes belonging to the same community compared to a $k$-core decomposition, and yields a more insightful view of how the network is organized. Indeed, if we know that a node is $k$-dense, then we immediately have richer information about the local structure of the node's neighborhood. We know not only that the node has degree at least $k-1$ and so do its $k-1$ neighbors, but also that with each of these neighbors the node shares at least $k-2$ common neighbors, forming a highly clustered structure of strongly overlapping triangles. As shown in~\cite{POPSIM}, this clustering indicates strong similarity between the involved ASes.

Compared to CPM, the $k$-dense decompositions is less sensitive to noise~\cite{MYPT}. From the computational perspective, while the $k$-dense decomposition is slightly more complex than the $k$-core decomposition, it is much less complex than CPM.

\subsection{Definitions}

The $k$-dense decomposition is a recursive graph decomposition \cite{KDENSE}, based on edge multiplicity \cite{EMULTI,NEMULTI}. The multiplicity $m_{G}(i,j)$ of edge $(i,j)$ in graph $G$ is the number of triangles in $G$ containing the edge, or equivalently, the number of common neighbors of connected nodes $i$ and $j$, see Figure \ref{fig:kdef}. By definition, the $k$-dense subgraph $H_k$ of graph $G$ is the subgraph induced by all the links with multiplicity larger or equal to $k-2$ {\em in the subgraph}:
\begin{equation}
{m}_{H_k}(i,j) \geq k-2.
\label{eq:kdense}
\end{equation}
This subgraph can be obtained from $G$ by iterative pruning all the links with multiplicity smaller than $k-2$. All the links in $H_k$ have multiplicity equal or greater than $k-2$ in $G$ as well, ${m}_{H_k}(i,j) \geq k-2 \Rightarrow {m}_{G}(i,j) \geq k-2$, but the converse is not generally true. That is, links with ${m}_{G}(i,j) \geq k-2$ are only {\em candidate} $H_{k}$ links.

\begin{figure*}[t]
\centerline{\subfloat[]{\includegraphics[width=0.3\textwidth]{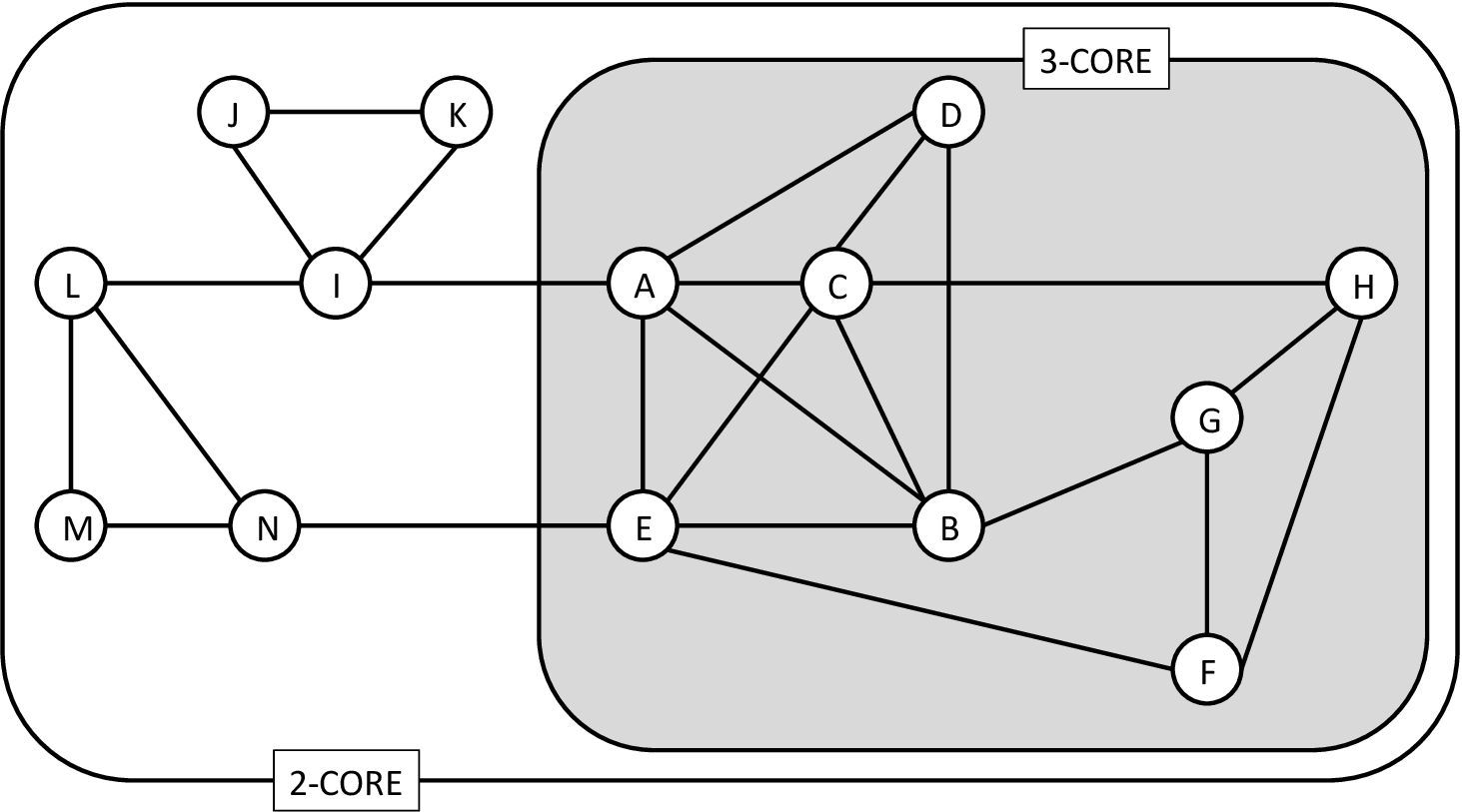}%
\label{subfig:excore}}
\hfil
\subfloat[]{\includegraphics[width=0.3\textwidth]{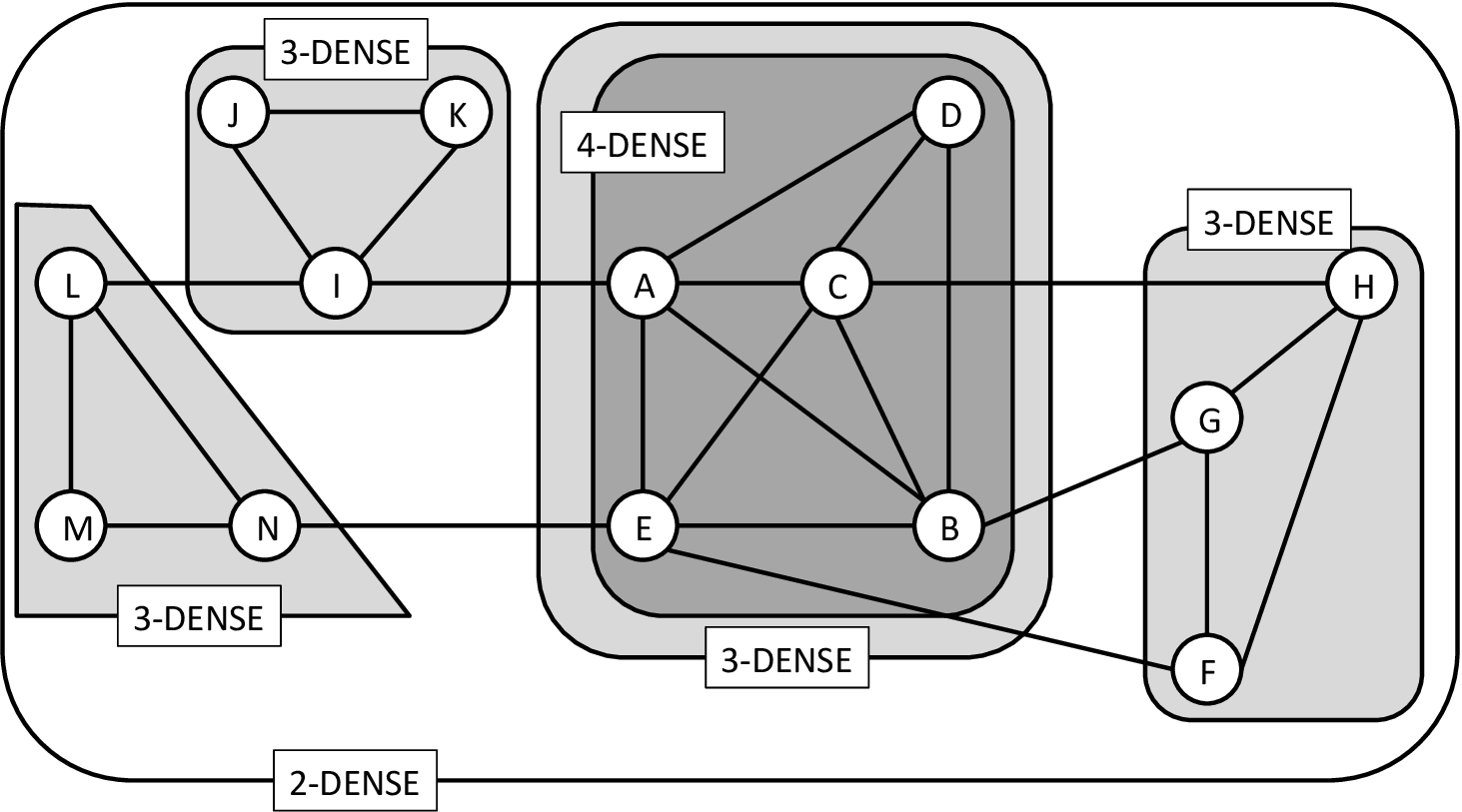}%
\label{subfig:exdense}}
\hfil
\subfloat[]{\includegraphics[width=0.3\textwidth]{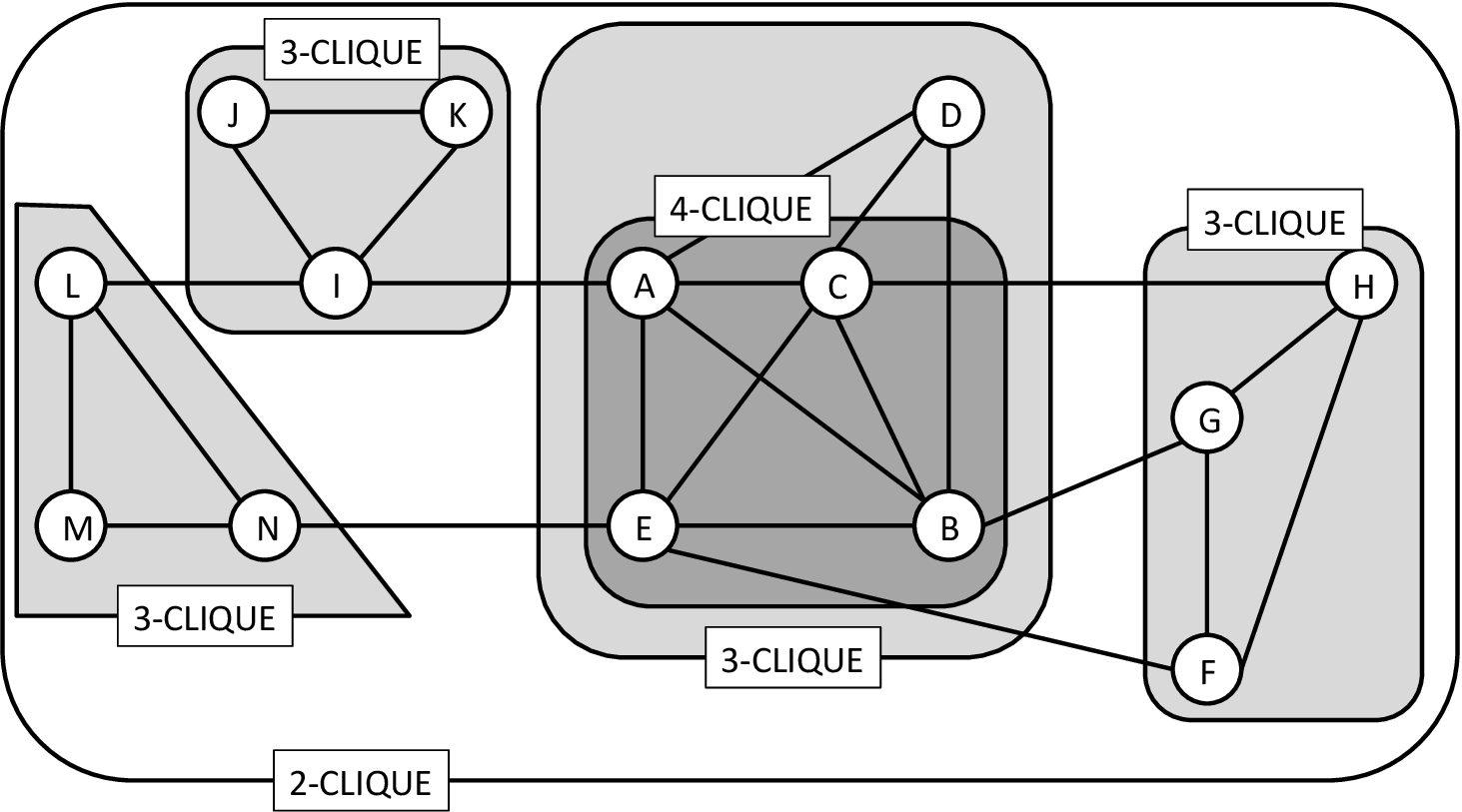}%
\label{subfig:exclique}}}
\caption Decompositions of a sample graph by the $k$-core~\protect\subref{subfig:excore}, $k$-dense~\protect\subref{subfig:exdense}, and $k$-cliques~(CPM)~\protect\subref{subfig:exclique} methods.
\label{fig:kcomparison}
\end{figure*}

The \textit{k}-dense decomposition of $G$ is a set of nested subgraphs $H_{k_{MAX}}\subset\ldots\subset H_{k+1}\subset H_k\subset\ldots\subset G$, where $H_{k_{MAX}}$ is the smallest and densest non-empty sub-graph.  An example of $k$-dense decomposition is shown in Figure \ref{subfig:exdense}.

A link is said to have the \textit{k-dense-index} equal to $k^*$ if it belongs to $H_{k^*}$ but not to $H_{k^*+1}$. The set of all the links with the \textit{k}-dense-index equal to $k^*$ is called the $k^*$-dense-shell.

A node is said to have the \textit{k-dense-index} equal to $k^*$ if $k^*$ is the maximum \textit{k}-dense-index of its incident links. The set of all the nodes with \textit{k}-dense-index equal to $k^*$ is called the $k^*$-dense-set.

\section{Data}
\label{sec:data}

\subsection{Description}

\begin{table*}[t]
\centering
\caption{ Number of AS nodes, AS links, and the values of $k_{MAX}$.}
\label{tab:growth}
\vspace{0.3cm}
\begin{tabular}{cccccccccc}
 & 2004 & 2005 & 2006 & 2007 & 2008 & 2009 & 2010 & 2011 & 2012 \\
\hline
\hline
Nodes & 17,858 & 20,486 & 23,044 & 26,101 &  29,042 & 32,379 &  35,583 & 38,888 &  42,419 \\
Links & 50,326 & 59,382 & 66,781 & 79,931  & 90,588 & 102,362  & 113,846  & 127,558 & 146,271 \\
$k_{MAX}$ & 29 & 33 & 32 & 34  & 40 & 37  & 39  & 42 & 48 \\
\hline
\end{tabular}
\end{table*}

We use the AS-level topology data from the UCLA Computer Science Department's Internet Research Lab (IRL) \cite{QUANT,IRL}. We collect yearly Internet snapshots dated from May 2004 to May 2012. Specifically, for each year, we download the data corresponding to May 31st, and then discard all the links with the \textit{last seen} attribute older than May 1st. Table \ref{tab:growth} reports the numbers of AS nodes and links in each snapshot.

We emphasize that BGP and traceroute-based data provide incomplete and biased views of the real Internet AS-level topology. For example, many connections between leaf (low-degree) ASes are hidden from monitors located in hub (high-degree) ASes~\cite{FEEDER}. Changing numbers of monitors introduce various artifacts and aberrations in the observed topologies. Since only the best paths are typically announced and used for routing, the observed views are highly incomplete. Specific to traceroute measurements, there are many non-responding ASes, especially leaf ASes comprising a majority of Internet ASes. There are also many issues with mapping IP addresses to AS numbers, including IP addresses for which there are multiple or no mapping ASes, and many other vagaries \cite{MIDAR}.

Our decision to use the IRL BGP data was motivated primarily by the observation that the number of ASes in this dataset is relatively consistent with the number of the Advertised AS Count growth in the CIDR report \cite{CIDR}.

\subsection{Incompleteness}
\label{sec:incompleteness}
While these Internet topology measurements are known to be incomplete, many studies have demonstrated that a vast majority of missing links are peering connections \cite{ASREL,LOL,ELUSIVEGT,LARGEIXP,FEEDER}. The reason for this is that
a topology measurement monitor located in AS~$X$ can observe only $X$'s peering connections and the peering connections between ASes that are above~$X$ in the customer-provider AS hierarchy~\cite{ASREL}. As a consequence, peering connections between large providers tend to be detected by topology measurements with high probability, while peering links between small ASes are mostly missing.

Based on these observations, we believe that a $k$-dense analysis of the complete topology would differ from the analysis below as follows:
\begin{itemize}
\item the $k_{MAX}$ index would probably be higher,
\item some ASes would have higher $k$-dense indices,
\item there would probably be a larger number of small $k$-dense communities disconnected from the giant $k$-dense components ($H_k$s consist of one connected component for most values of~$k$~\cite{MYPT}).
\end{itemize}
At the same time, a considerable percentage of low-degree ASes would unlikely be affected by this incompleteness because most ASes whose business is not related to Internet operations connect only to their providers via \textit{observable} customer-provider links.

\section{Internet's $k$-dense decomposition}
\label{sec:analysis}

Treating the data described in Section \ref{sec:data} as a sorted set of AS-level graphs ordered by year, we report in this Section the results of the statistical analysis of Internet's $k$-dense decomposition and its evolution over time. The interpretation of these results is deferred to Section \ref{sec:discussion}.

\subsection{Basic trends}\label{sec:trends}

\begin{figure}[t]
\centering
\includegraphics[width=0.5\textwidth]{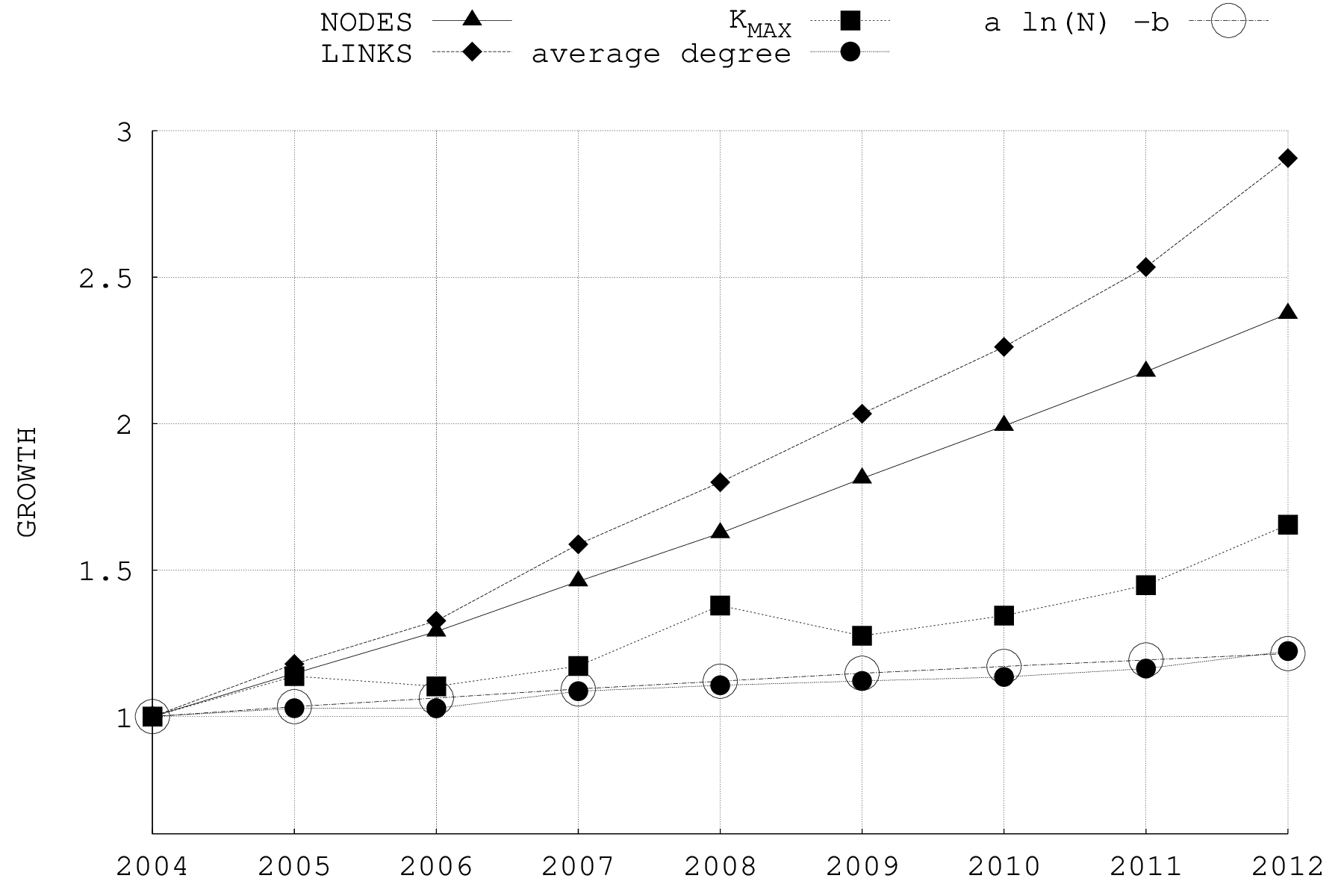}
\caption{Internet growth in terms of numbers of nodes, links, ${k}_{MAX}$-dense-index and average degree. The black triangles show ${N(t)}/{N(t_0)}$, the number of nodes in the graph at time $t$ divided by the number of nodes at time $t_0 = 2004$, $N(t_0)=17,858$.  The black rhombuses show the same ratio for the number of links, ${M(t)}/{M(t_0)}$, $M(t_0)=50,326$. The black squares are ${{k_{MAX}}(t)}/{{k_{MAX}}(t_0)}$, $k_{MAX}(t_0) = 29$. The black circles show the average degree of the graph at time $t$ divided by the average degree at time $t_0 = 2004$, $\overline{k}(t_0)=5.64$. The average degree grows logarithmically with the Internet size, $\bar{k} \approx a \ln N - b$, with $a=1.3$ and $b=7.5$. The empty circles are $(a \ln N(t) - b)/(a \ln N(t_0) - b)$, showing the quality of this approximation.}
\label{fig:internetgrowth}
\end{figure}

Similar to other studies, e.g.~\cite{ASEVOLUTION,TWELVE}, we have not observed and do not report any significant changes in the basic graph properties---degree distribution, degree correlations, clustering, betweenness, and shortest path distributions---even though the graph has grown significantly, Table \ref{tab:growth} and Figure \ref{fig:internetgrowth}.
We see that the number of links $M$ has been growing faster than the number of nodes $N$, meaning that the average degree $\bar{k}=2M/N$ has been increasing. This increase appears to be a logarithmic function of $N$,  $\bar{k} \approx a \ln N - b$, with $a=1.3$ and $b=7.5$, in agreement with \cite{POPSIM,UNIVERSE}.
The increasing trend of ${k}_{MAX}$ growing from 29 in 2004 to 48 in 2012 indicates that more densely connected parts in the Internet have been forming in the course of its growth.

\subsection{$k$-dense normalization}\label{sec:normalization}

For each snapshot we first compute the number of nodes and links with a given \textit{k}-dense-index, i.e.\ the number of nodes in each \textit{k}-dense-set and the number of links in each \textit{k}-dense-shell. Since the graph in the snapshots and its $k_{MAX}$ are growing, to be able to properly compare the nine snapshots we next perform the following normalization, mapping $k$-dense-indices and corresponding numbers of nodes and links to fractions with values between $0$ and $1$:
\begin{itemize}
\item \textit{x-axis normalization}: map each index \textit{k} in each snapshot to what we call the {\em \textit{k}-dense-index fraction}:
\begin{equation}
x =\frac{k - k_{MIN}}{k_{MAX} - k_{MIN}},
\label{eq:kdf}
\end{equation}
where $k_{MIN}$ and $k_{MAX}$ are the minimum and maximum values of the $k$-dense-index in the snapshot;
\item \textit{y-axis normalization}: divide the corresponding number of nodes or links by the total number of nodes or links in the snapshot.
\end{itemize}

In all the considered snapshots there are no nodes of degree $0$, so that $k_{MIN}=2$ for links and nodes in all the snapshots. This means that $x=0$ corresponds to the $k$-dense-index equal to $2$, while $x=1$ corresponds to the $k$-dense-index equal to $k_{MAX}$, which grows with time. Upon such normalization it becomes difficult to read off the absolute values of $k$-dense-indices, nodes, or links from the resulting plots, but this normalization is helpful to see if there are any size- and time-invariant statistical trends.

\subsection{$k$-dense decomposition}\label{sec:kdd}

\begin{figure*}[t]
\centerline{\subfloat[]{\includegraphics[width=0.32\textwidth]{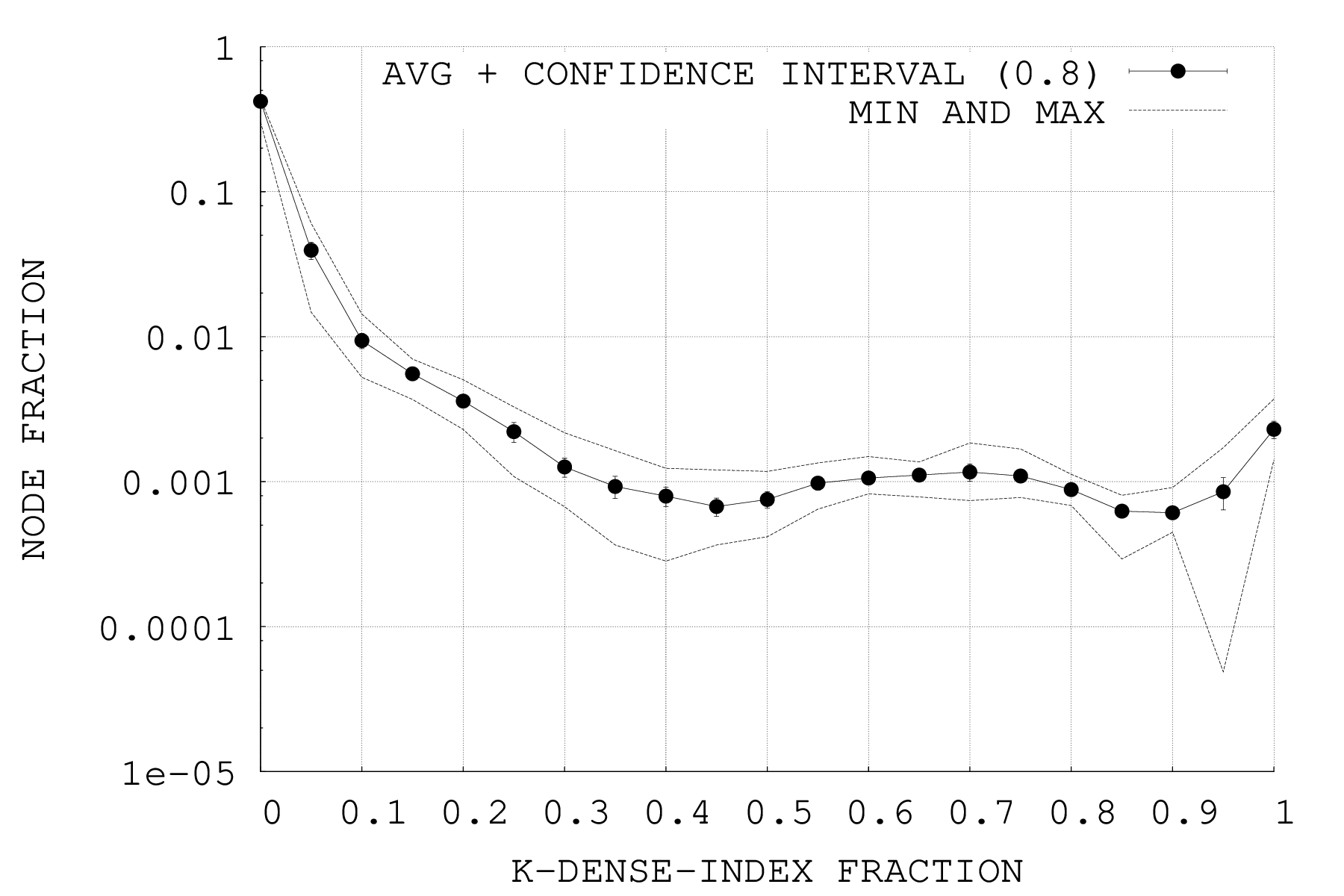}%
\label{subfig:denseass}}
\hfil
\subfloat[]{\includegraphics[width=0.32\textwidth]{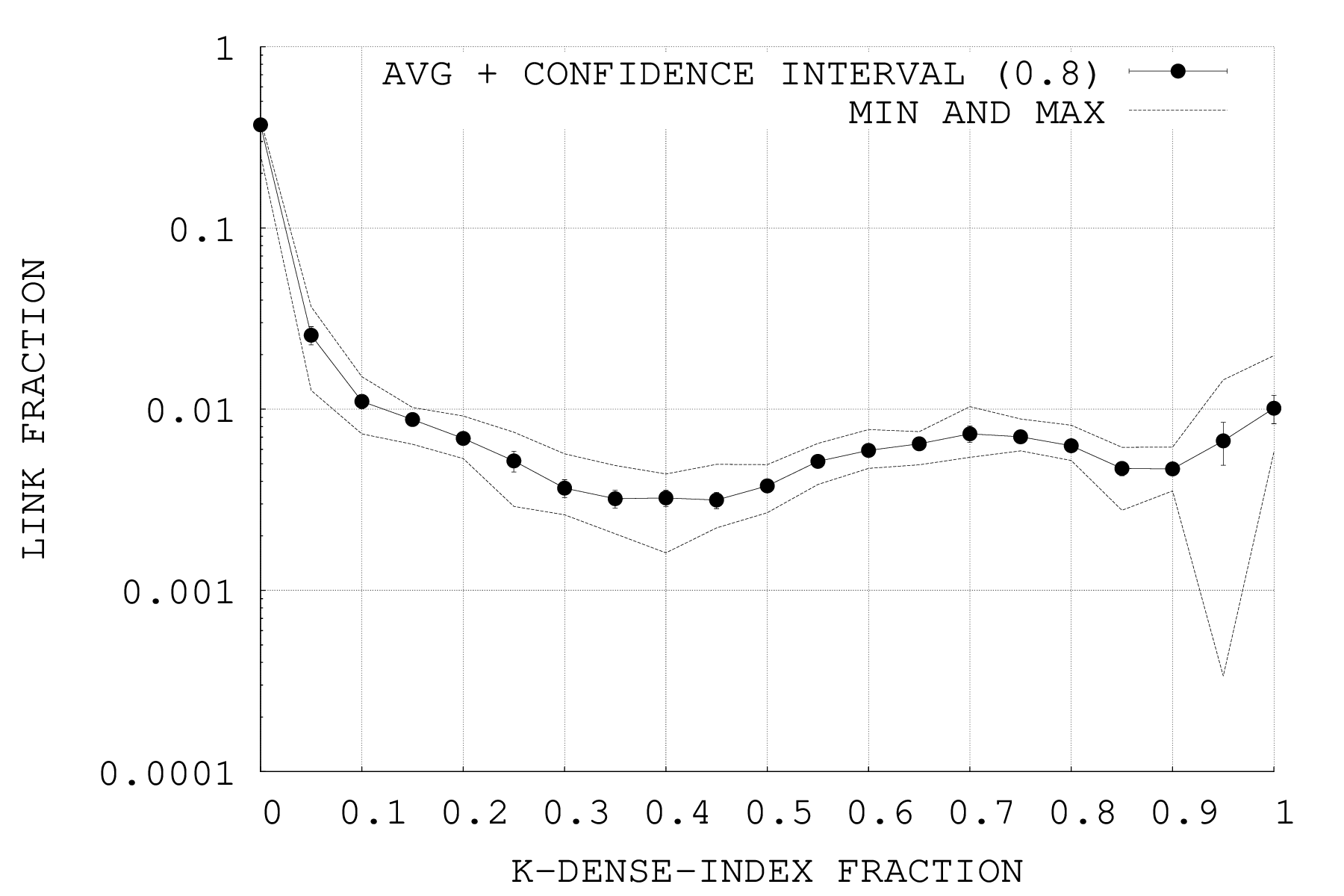}%
\label{subfig:denseconnections}}
\hfil
\subfloat[]{\includegraphics[width=0.32\textwidth]{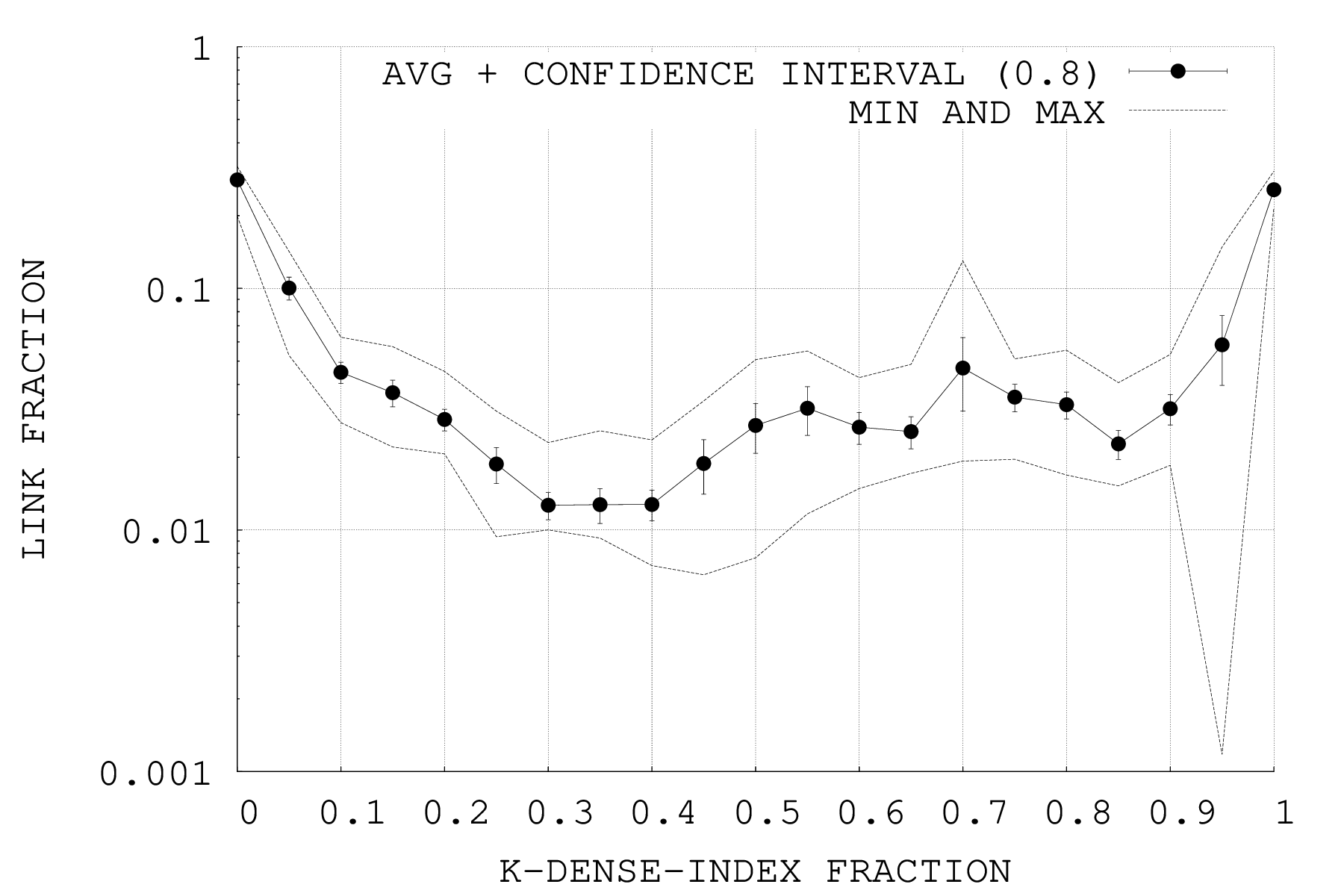}%
\label{subfig:internetkdsconnectionsbn}}}
\centerline{
\subfloat[]{\includegraphics[width=0.32\textwidth]{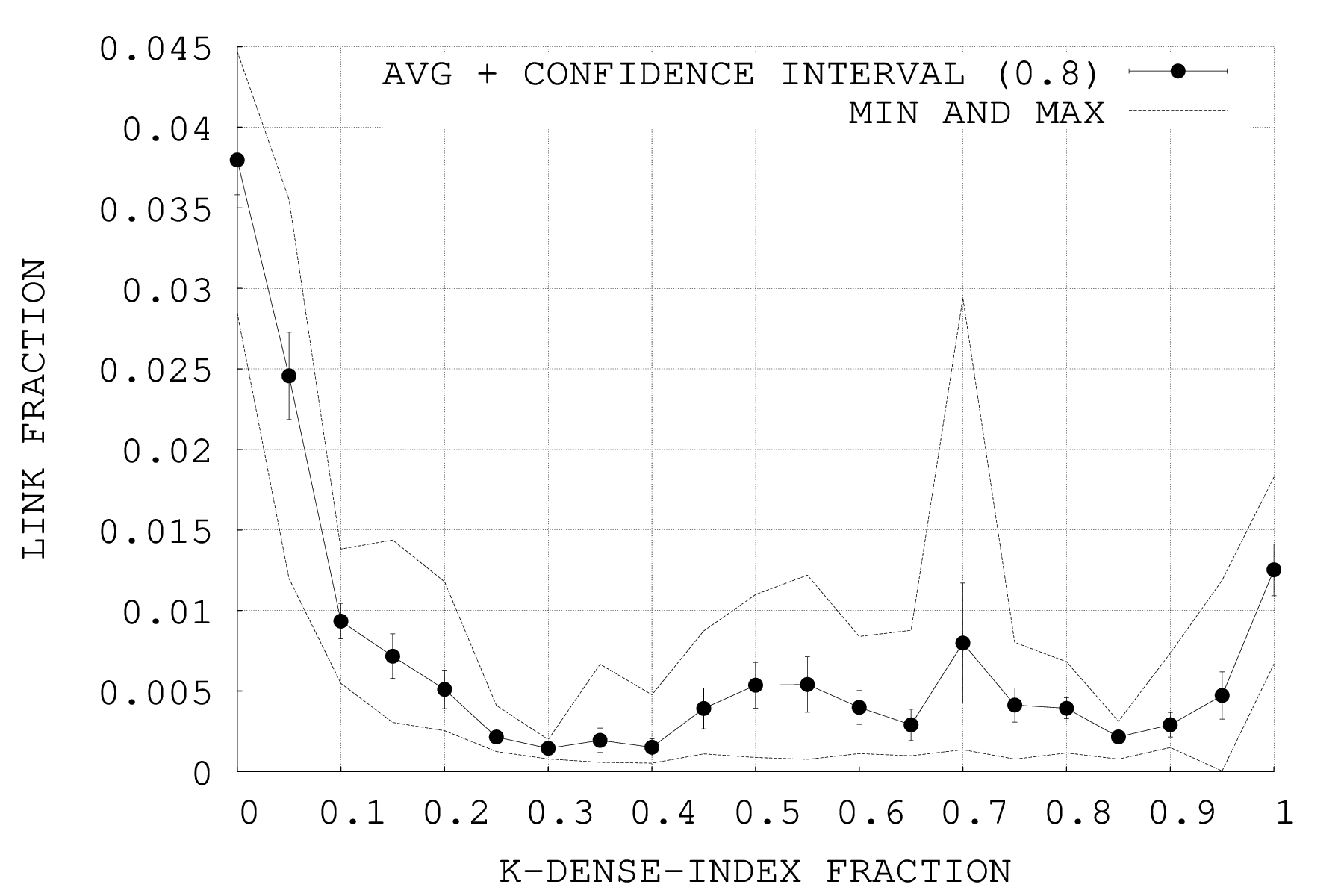}
\label{subfig:2connections}}
\hfil
\subfloat[]{\includegraphics[width=0.32\textwidth]{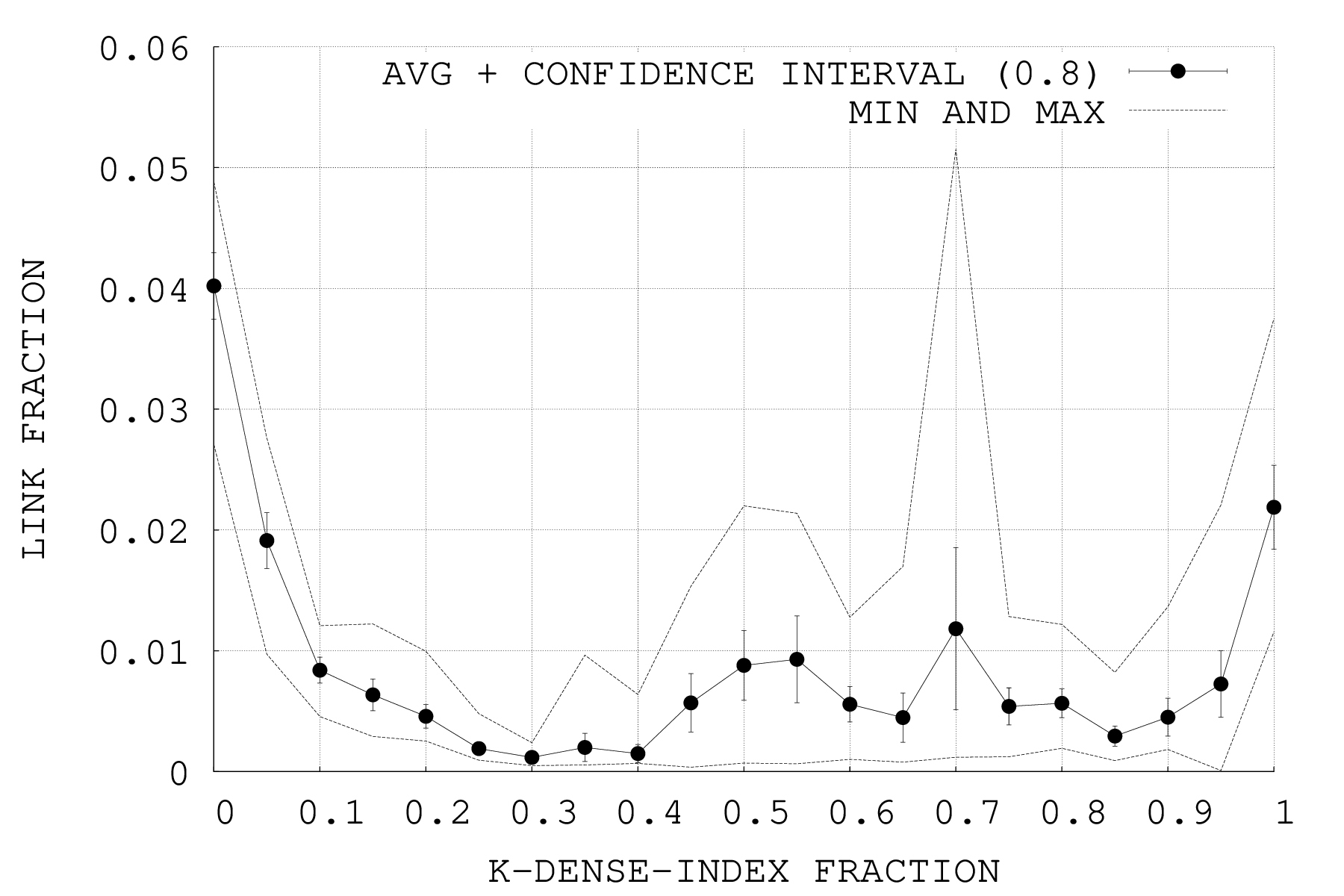}
\label{subfig:3connections}}
\hfil
\subfloat[]{\includegraphics[width=0.32\textwidth]{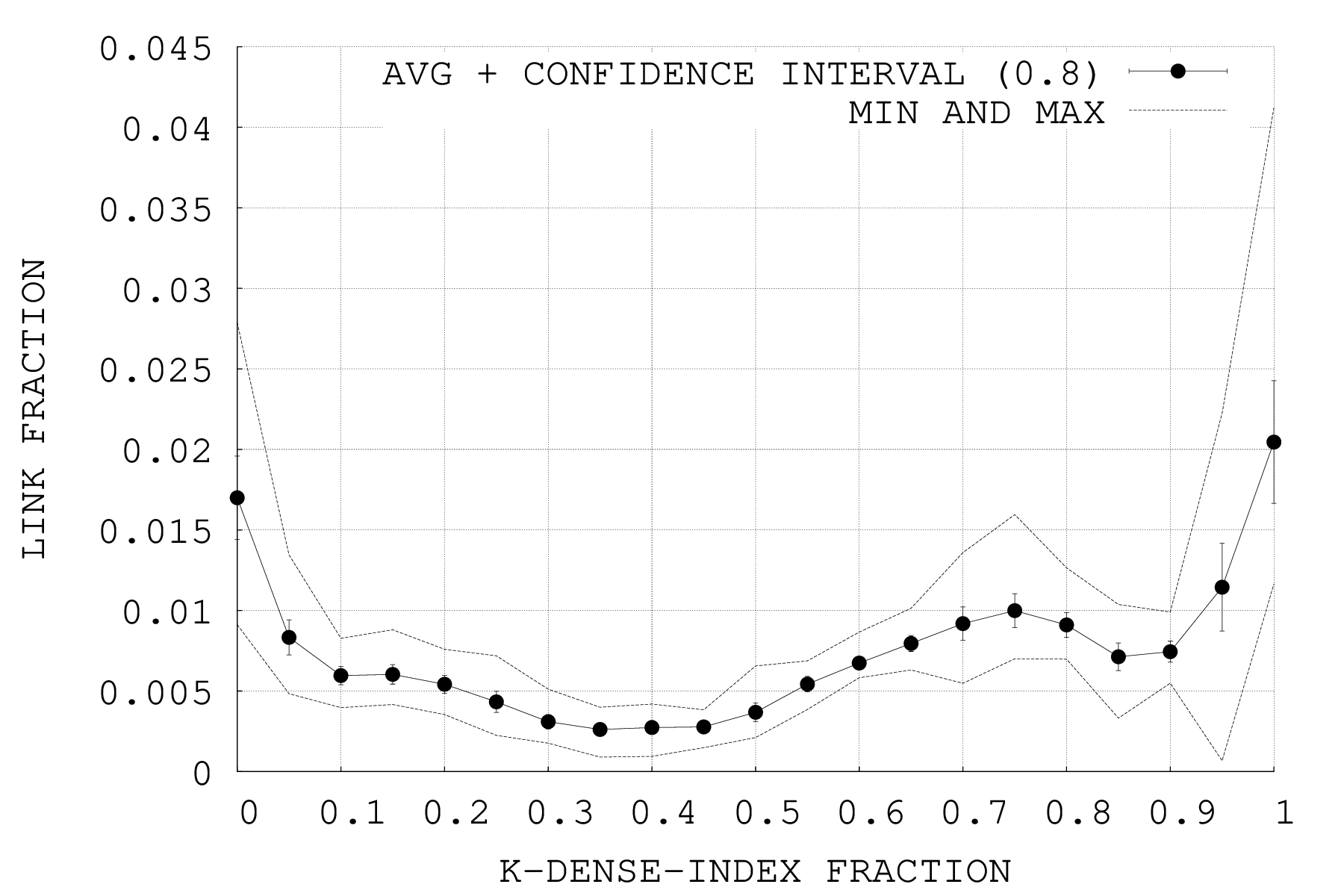}
\label{subfig:kmaxconnections}}}
\caption{Normalized $k$-dense decomposition of the Internet. The $x$-axes are linearly binned with the bin size equal to $0.05$, and each plot shows the average (black dots and solid lines), $80\%$ percentile (vertical bars), and minima and maxima (dashed lines) of the corresponding values computed across the nine historical snapshots of the Internet. \protect\subref{subfig:denseass} is the fraction of nodes with a given \textit{k}-dense-index, i.e.~the number of nodes in the \textit{k}-dense-set divided by the total number of nodes in the corresponding snapshot. \protect\subref{subfig:denseconnections} is the fraction of links with a given \textit{k}-dense-index, i.e.~the number of links in the \textit{k}-dense-shell divided by the total number of links in the corresponding snapshot. \protect\subref{subfig:internetkdsconnectionsbn} is the fraction of links attached to a given $k$-dense-set, i.e.~the number of links whose one or two ends are attached to nodes in a given \textit{k}-dense-set, divided by the total number of links in the corresponding snapshot.
\protect\subref{subfig:2connections} is the number of links with one end attached to a $2$-dense-set node and with the other end attached to a \textit{k}-dense-set node, divided by the number of links attached to the $2$-dense-set. \protect\subref{subfig:3connections} and \protect\subref{subfig:kmaxconnections} show the corresponding fractions for links attached to $3$- and $k_{MAX}$-dense-sets.}
\end{figure*}

In Figures \ref{subfig:denseass} and \ref{subfig:denseconnections} we report the normalized fractions of nodes and links for each \textit{k}-dense-index, averaged over the nine historical snapshots. These average values are highly representative for each snapshot from 2004 to 2012, in the sense that the distributions of the actual values across the snapshots are quite narrow, as indicated by the $80\%$ percentile and min/max deviations from the averages in the figures. In other words, the shown normalized statistics are fairly stable over time, i.e.~time- and size-invariant. In particular, a vast majority of nodes and links in all the snapshots have low $k$-dense-indices, not belonging to any dense communities.

Similar to the joint degree distribution, showing how nodes of different degree interconnect~\cite{DKGRAPHS}, we next focus on how different $k$-dense-sets interconnect.
In Figure \ref{subfig:internetkdsconnectionsbn} we report the normalized fraction of links attached to nodes in a given $k$-dense-set. Similarly to Figures \ref{subfig:denseass} and \ref{subfig:denseconnections}, we observe that this statistics is also stable over time. We also observe
that there are two classes of \textit{k}-dense-sets to which a vast majority of all links are attached---those with the smallest and largest \textit{k}-dense-index fractions close to $0$ and $1$. Upon examination of the original data we find that the left peak in Figure \ref{subfig:internetkdsconnectionsbn} is formed by the links attached to the $2$- and $3$-dense sets, while the right peak is due to the links attached to the $k_{MAX}$-dense set. This observation motivates us to further focus on the links attached to these three sets, and show where their other ends go in Figures \ref{subfig:2connections}-\ref{subfig:kmaxconnections}.
We see that $2$- and $3$-dense-attached links go mostly to other low-dense-index nodes and to the nodes in the $k_{MAX}$-dense-set, and that ${k}_{MAX}$-dense nodes direct a considerable percentage of their connections not only to nodes in low-dense-sets, but also to nodes in densely-connected parts of the network. We also notice that all the considered normalized statistics appear to be quite stable over time, as indicated by the narrow $80\%$ percentiles.

Given that a vast majority of links are attached to nodes in $2$-, $3$-, or $k_{MAX}$-dense sets, we further characterize these sets in Table \ref{tab:avg23k}. The significant percentages of links attached to $2$- and $3$-dense sets are not surprising in view of that a majority of nodes belong to these sets, Figure \ref{subfig:denseass}. We see in Table \ref{tab:avg23k} that these peripheral sets are populated with nodes of low average degree and betweenness centrality of the order of the number of nodes $N$ in the graph.
The ${k}_{MAX}$-dense-set, on the other hand, consists of a small number of nodes with high average degree and betweenness of the order of $N^2$. That is, these nodes have a central position in the network, they are a key element shaping the overall connectivity, with many links attached to them, Figure \ref{subfig:internetkdsconnectionsbn}, which motivates us to further analyze the structure of this $k_{MAX}$-dense subgraph. However before doing so in Section \ref{sec:kd}, we first answer the following two natural questions:
\begin{enumerate}
\item Does the degree of a node fully define its $k$-dense-index, making this index a repetitive statistics providing no new information compared to node degree?
\item Do random graphs having the same degree distribution or degree correlations as the Internet, fully reproduce all its $k$-dense properties, making them statistically insignificant, that is, casting them as statistical consequences of that this graph has the observed (joint) degree distribution?
\end{enumerate}

\begin{table}
\centering
\caption{Average properties of nodes in the $2$-, $3$-, and ${k}_{MAX}$-dense-sets: average degree $\overline{k}$, average clustering coefficient $\overline{c}$, and average betweenness centrality $\overline{b}$.}
\label{tab:avg23k}
\begin{tabular}{cccc}
 & $\overline{k}$ & $\overline{c}$ & $\overline{b}$ \\
\hline
\hline
\textit{$2$-dense-set}					&		1.643 		& 0 			& $2.1 \cdot 10^3$ 	\\
\textit{$3$-dense-set}					& 		3.161 		& 0.746	& $9.3 \cdot 10^3$ 	\\
${k}_{MAX}$\textit{-dense-set} 	& 		403.8 	& 0.194 	& $3.0 \cdot 10^6$ 		\\
\hline
\end{tabular}
\end{table}

\subsection{$k$-dense-index versus node degree}\label{sec:density.vs.degree}
\begin{figure*}[t]
\subfloat[]{\includegraphics[width=0.45\textwidth]{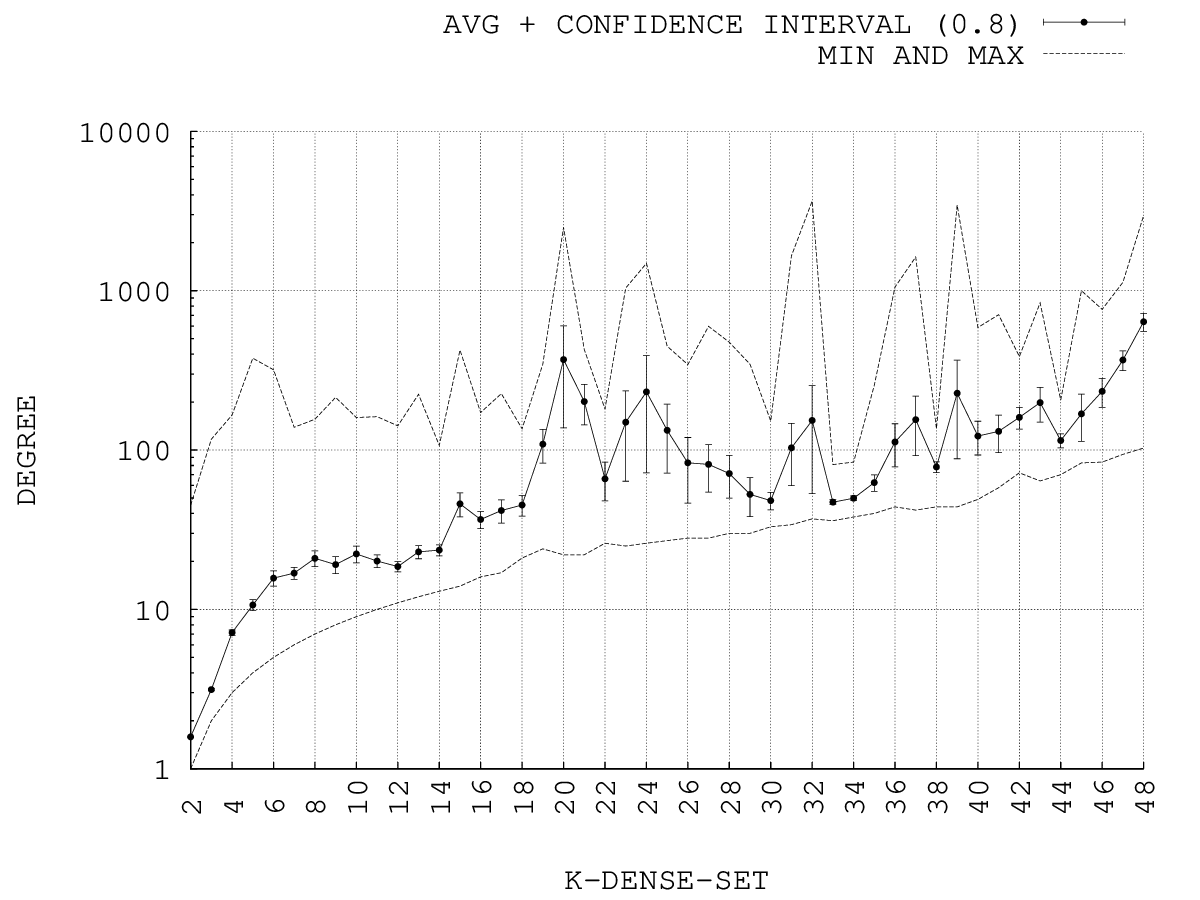}%
\label{subfig:kd2012}}
\hfil
\subfloat[]{\includegraphics[width=0.45\textwidth]{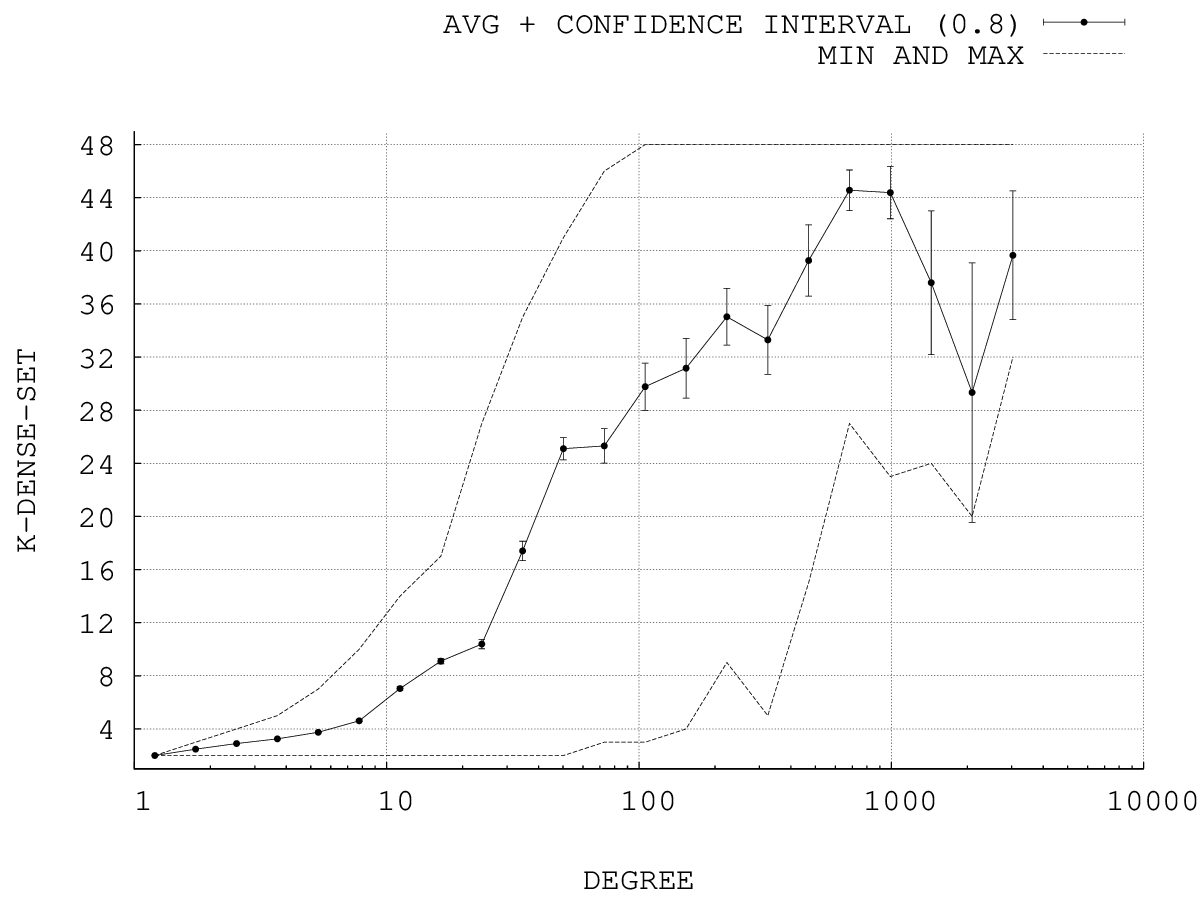}%
\label{subfig:dk2012}}
\caption{The degrees of nodes versus their $k$-dense-indices in the 2012 Internet snapshot. The figures show the averages, $80\%$ percentiles, and min/max values for each distribution:
\protect\subref{subfig:kd2012} distributions of node degrees in $k$-dense-sets, and 
\protect\subref{subfig:dk2012} distributions of $k$-dense-indices of nodes with degrees falling within a given range obtained by logbinning the degree values.}
\label{fig:dd2012}
\end{figure*}

By definition, a node with $k$-dense-index equal to~$k$ has degree greater or equal to $k-1$, while a node with degree $k$ has a $k$-dense-index smaller or equal to $k+1$. These constraints do not preclude strong relative fluctuations between these two properties within the allowed bounds. For example, a node with a high degree can be the center of a hub-and-spoke configuration, and have a lower $k$-dense-index than a node with a lower degree belonging to a small clique.
%If all nodes of the same degree have the same $k$-index, then one determines the other, so that the second statistics has no independent value.
In Figure~\ref{fig:dd2012} we show the relation between the $k$-dense-indices of nodes and their degrees in the 2012 snapshot. As expected the average degree of nodes in a given $k$-index-set as a function of $k$ exhibits a growing trend, however the fluctuations are high. In particular, the degrees of nodes in the $2$-dense-set vary from 2 to 45, while the degree of nodes in the $k_{MAX}$-dense-set ($k_{MAX}=48$) vary from 103 to 2964. The other way around, nodes of degree $100$, for example, have $k$-indices ranging from 12 to 45, and the node with maximum degree 174 (Cogent Communications inc., $k=3655$) has the \textit{k}-dense-index equal to $32 < k_{MAX}$. We conclude that the $k$-index is not fully determined by the node degree, an observation that gains further support and Internet-specific explanation in Section~\ref{sec:discussion}.

\subsection{Statistical significance of the $k$-dense properties}\label{sec:dk-analysis}

\begin{figure}
\centering
\includegraphics[width=0.5\textwidth]{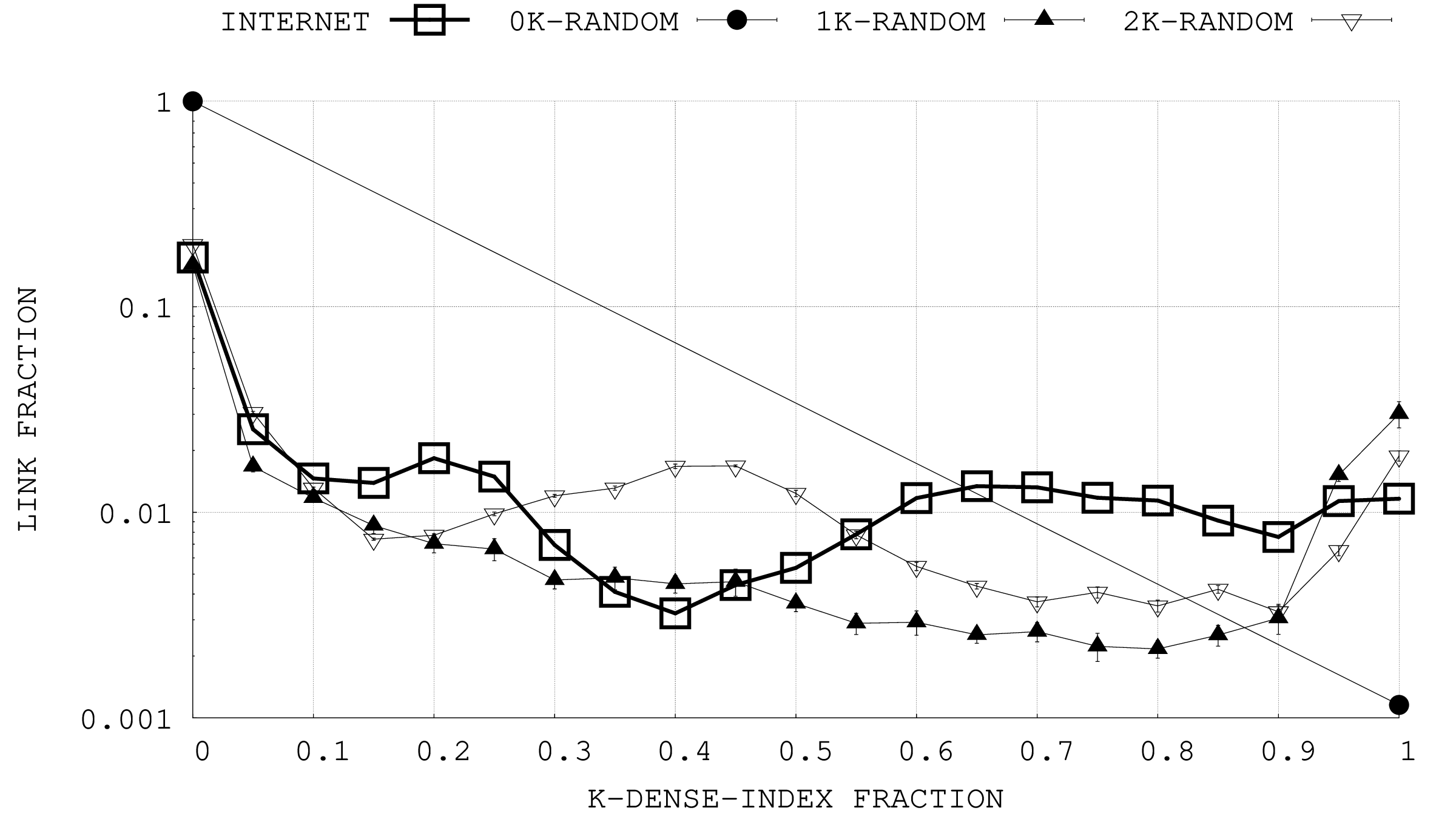}
\caption{Normalized $k$-dense decomposition of links in Internet 2012 and its $dK$-randomizations for $d=0,1,2$. The figure is similar to Figure \ref{subfig:denseconnections}, and shows the averages (data points and solid lines) and $80\%$ percentiles (vertical bars) of the distributions of $k$-dense link fractions in $dK$-random graph instances for $d=0,1,2$.}
\label{fig:rand_comp}
\end{figure}

Even though the node degree does not fully determine the node $k$-index, it may still be the case that random graphs having the same degree distribution as the Internet, also have exactly the same $k$-dense properties, meaning that these properties are nothing but statistical artifacts, as it turned out to be the case with the rich club connectivity in the Internet \cite{RCLUB}. To check if a similar story applies to the $k$-dense properties, we construct $dK$-random graphs~\cite{DKGRAPHS} for $d=0,1,2$ as described in Section \ref{sec:kd}. We generate 10 realizations for each \textit{d}. These graphs are random graphs with the same average degree, degree distribution, or joint degree distribution as in the Internet 2012 snapshot. In Table \ref{tab:kmax} we report the distributions of the $k_{MAX}$ values in these random graphs for each $d$, while in Figure \ref{fig:rand_comp} we show the $k$-dense link decompositions of these graphs, and juxtapose them against the Internet's. We observe that neither degree distribution nor joint degree distribution fully reproduce the $k$-dense properties of the Internet, meaning that these properties have their own statistical significance, which is another observation that finds a natural Internet-specific explanation in Section~\ref{sec:discussion}. Yet another property that finds an explanation rooted in Internet specifics in that section is an unexpectedly simple structure of the $k_{MAX}$-index core that we analyze next.

\begin{table}
\centering
\caption{$k_{MAX}$-dense-index in Internet 2012 and its $dK$-randomizations for $d=0,1,2$. The table shows the averages $\bar{k}_{MAX}$ and standard deviations $\sigma$ of the distributions of the $k_{MAX}$-index values in $dK$-random graph instances for $d=0,1,2$.}
\label{tab:kmax}
\begin{tabular}{ccc}
 & $\bar{k}_{MAX}$ & $\sigma$ \\
\hline
\hline
Internet 	& 48 	& - 	\\
$0K$-random 	& 3		& 	0		\\
$1K$-random	& 68		& 	4.05		\\
$2K$-random	& 44.3	& 	0.46		\\
\hline
\end{tabular}
\end{table}

\subsection{Structure of the $k_{MAX}$-dense core}
\label{sec:kd}

\begin{table}
\centering
\caption{Basic properties of $k_{MAX}$-dense cores in the nine historical snapshots: $N$ the number of nodes; $M$ the number of links, $D=2M/[N(N-1)] \approx\bar{k}/N$ the graph density.}
\begin{tabular}{ccccc}
Year	&	${k}_{MAX}$	&	$N$	&	$M$	&	$D$ \\
\hline
\hline
2004 &	29	&	59	&	1,381	&	0.807 \\
2005 &	33	&	55	&	1,318	&	0.888 \\
2006 &	32	&	44	&	872		&	0.922 \\
2007 &	34	&	97	&	3,159	&	0.678 \\
2008 &	40	&	63	&	1,763	&	0.903 \\
2009 &	37	&	55	&	1,353	&	0.911 \\
2010 &	39	&	60	&	1,595	&	0.901 \\
2011 &	42	&	81	&	2,745	&	0.847 	\\
2012 &	48	&	60	&	1,703	&	0.962 \\
\hline
\end{tabular}
\label{tab:avgkmax}
\end{table}

Given the importance of $H_{k_{MAX}}$, the densest and innermost subgraph core, Section \ref{sec:kdd}, we next characterize its structure in more detail. In Table~\ref{tab:avgkmax} we show the basic properties of $H_{k_{MAX}}$ subgraphs in all the nine snapshots. We see that these subgraphs are quite small and dense (link density $D=1$ in complete graphs).

\begin{figure*}
\centering
\subfloat[]{\includegraphics[width=0.45\textwidth]{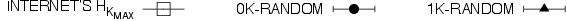}\label{subfig:key}} \\
\begin{tabular}{cc}
\subfloat[]{\includegraphics[width=0.45\textwidth]{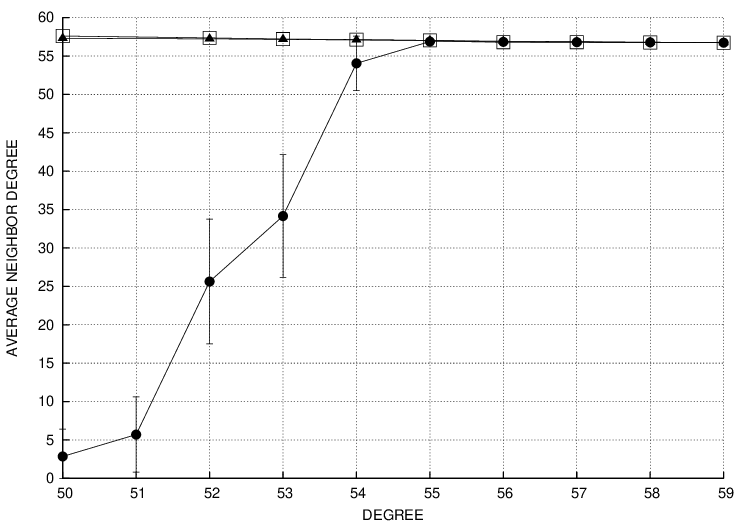}\label{subfig:2012knn}}
   & \subfloat[]{\includegraphics[width=0.45\textwidth]{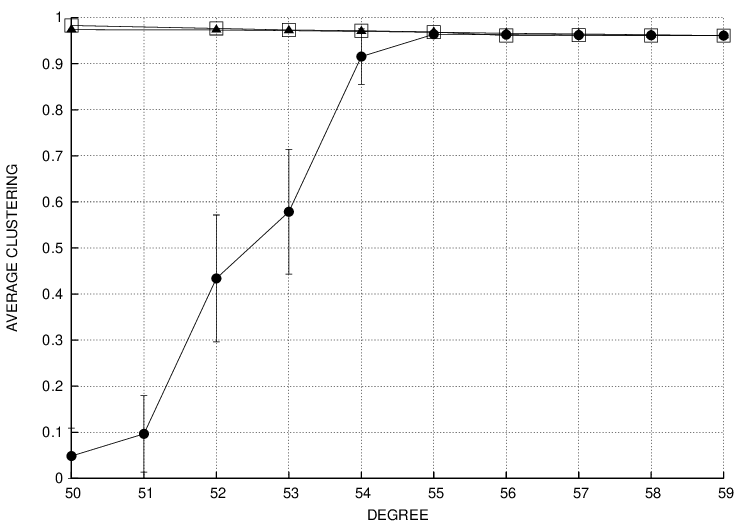}\label{subfig:2012cc}}\\
\subfloat[]{\includegraphics[width=0.45\textwidth]{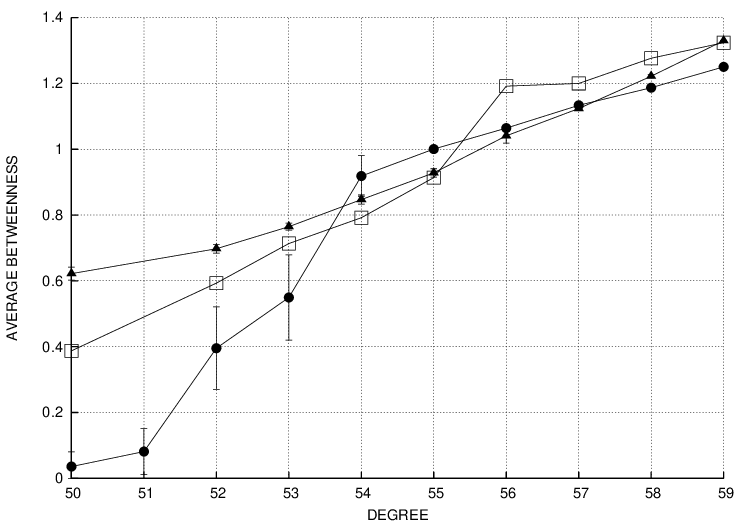}\label{subfig:2012bet}}
   & \subfloat[]{\includegraphics[width=0.45\textwidth]{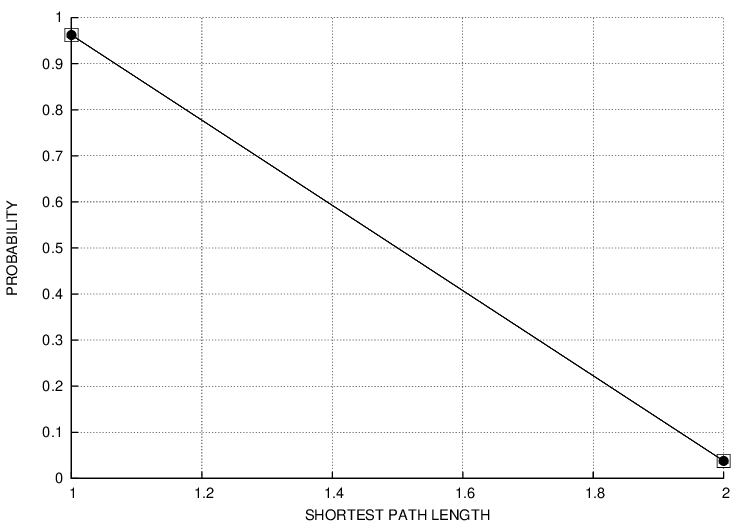}\label{subfig:2012path}}\\
   \subfloat[]{\includegraphics[width=0.45\textwidth]{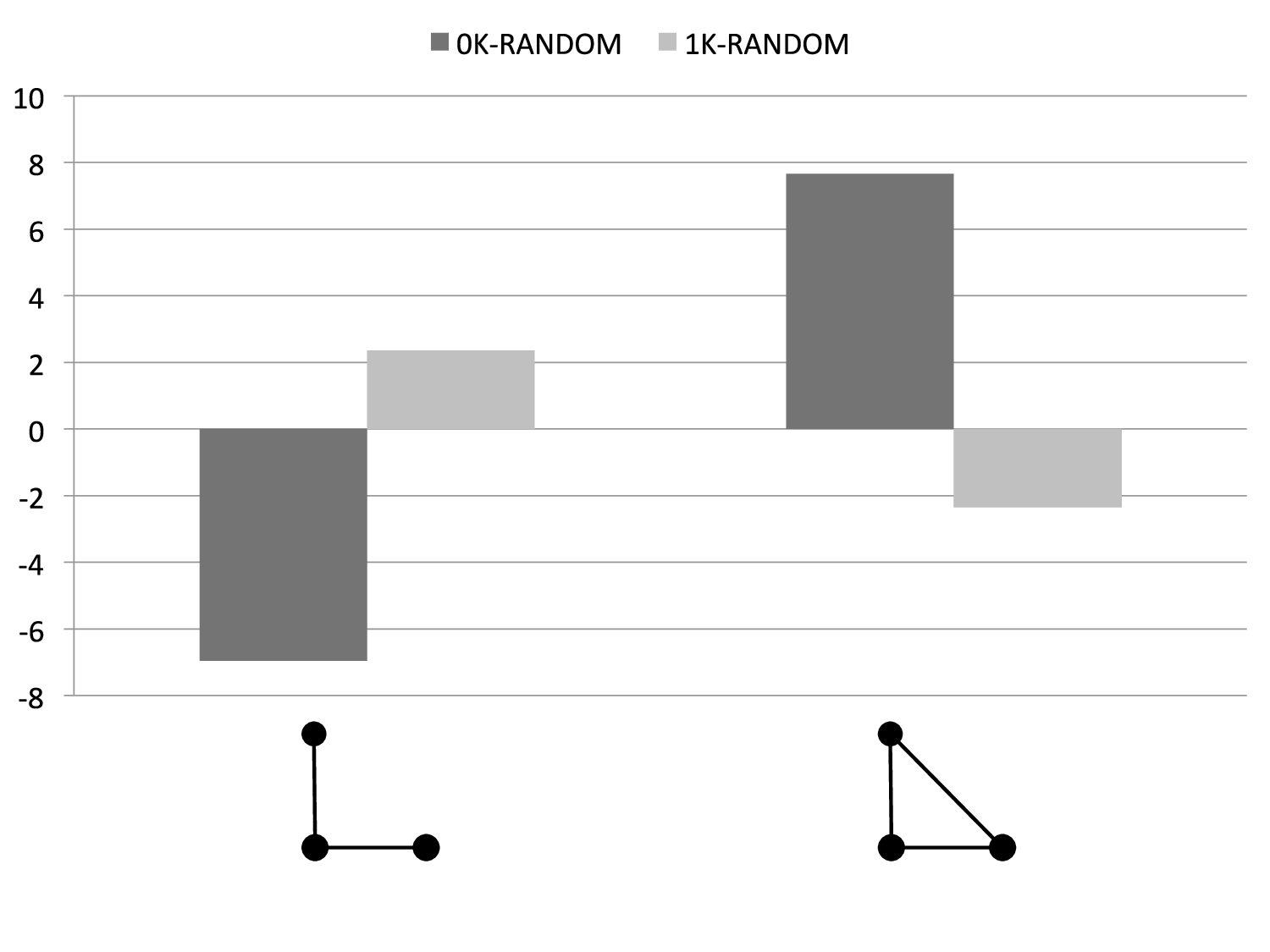}
\label{subfig:2012motif3}}
&\subfloat[]{\includegraphics[width=0.45\textwidth]{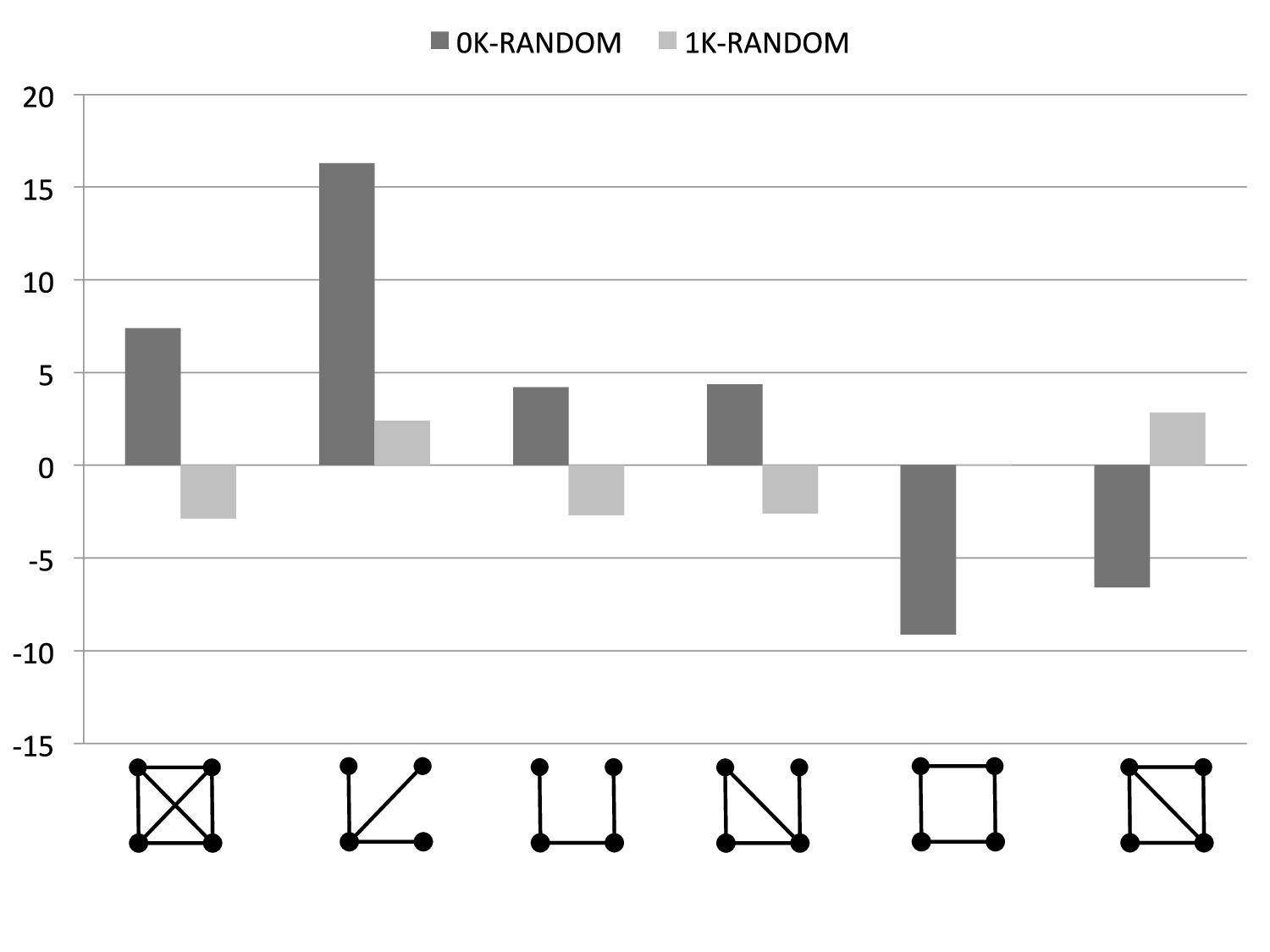}
\label{subfig:2012motif4}}
\end{tabular}
\caption{Basic structural properties of the $H_{k_{MAX}}$ core in Internet 2012 vs.\ its $dK$-randomizations for $d=0,1$. The figure shows the averages (points and solid lines) and $80\%$ percentiles computed across different $dK$-random graph instances. Plots \protect\subref{subfig:2012knn}, \protect\subref{subfig:2012cc}, and \protect\subref{subfig:2012bet} show the average neighbor degree, clustering, and betweenness of nodes of a given degree, while \protect\subref{subfig:2012path} shows the shortest path length distribution. Plots \protect\subref{subfig:2012motif3} and \protect\subref{subfig:2012motif4} show the $z$-scores $z$ for motifs of size $3$ and $4$, $z=(x-\mu)/\sigma$, where $x$ is a number of occurrences of a given subgraph in $H_{k_{MAX}}$, while $\mu$ and $\sigma$ are the mean and standard deviation of the distribution of the corresponding occurrences computed across different $dK$-random graph instances.}
\label{fig:2012avgdist}
\end{figure*}

Similar to the previous section, we next perform the standard $dK$-statistical analysis~\cite{DKGRAPHS} of these subgraphs. The procedure is as follows. First we focus on the latest 2012 snapshot, extract the $H_{k_{MAX}}$ subgraph from it, and treat this subgraph as a separate graph. In particular, node degrees are computed within $H_{k_{MAX}}$. Then we construct $20$ random graphs having exactly the same average degree as this graph, using the standard ${\mathcal G}_{N,M}$ (Erd\H{o}s-R\'{e}nyi \cite{ERMODEL}) construction procedure of throwing $M$ edges onto $N(N-1)/2$ node pairs uniformly at random. These graphs are called $0K$-random graphs. Then we construct $20$ random graphs having exactly the same degree sequence as $H_{k_{MAX}}$ using the fast generalized Havel-Hakimi algorithm from \cite{ZOLTAN}. This algorithm is guaranteed to always quickly succeed as soon as the degree sequence is graphical, i.e.\ realizable. The resulting graphs are called $1K$-random graphs. Then we compute the basic structural graph properties that in neither $0K$- nor $1K$-random graphs are guaranteed to be the same as in the original $H_{k_{MAX}}$ graph. The results are in Figure \ref{fig:2012avgdist}.

The average neighbor degree is not guaranteed to be captured by $1K$-random graphs. Only $2K$-random graphs, having exactly the same joint degree distribution as the original graph, guarantee to reproduce this degree correlation statistics. Yet we find that $1K$-random graphs have the same average neighbor degree as $H_{k_{MAX}}$. Clustering is guaranteed to be fully captured only by $3K$-random graphs, reproducing the frequencies of triangular subgraphs. Yet we find that $1K$-random graphs have the same clustering as $H_{k_{MAX}}$. The frequencies of motifs of size $3$ or $4$ are guaranteed to be fully captured only by $3K$- or $4K$-random graphs, the latter reproducing the frequencies of subgraphs of size $4$ by definition, but we find that there are no statistically significant deviations of the frequencies of these motifs in $1K$-random graphs from the corresponding counts in $H_{k_{MAX}}$. Finally the global betweenness or shortest path length distributions are not guaranteed to be captured by $dK$-random graphs with any $d<N$, but again we see that these global statistics are well reproduced in $1K$-random graphs, although not as well as the local statistics. At the same time none of the considered structural properties is closely approximated by $0K$-random graphs (except for the shortest path length distribution).

Collectively, these observations imply that even though the $H_{k_{MAX}}$ core is so dense (link density $D=0.962$) and close to being a complete graph that one could expect it to be well approximated by $0K$-random graphs, this expectation is actually wrong, but $1K$-random graphs capture closely the basic structural properties of this core.

We applied the same $dK$-analysis to the $H_{k_{MAX}}$ cores in all the other snapshots, and found them all to be $1K$-random as well. In other words, $1K$-randomness of the $H_{k_{MAX}}$ core is another time-invariant property.

\section{Interpretation of Internet's $k$-dense properties}
\label{sec:discussion}

In this section, we interpret and explain the statistical results from the previous section using specifics of the Internet, where nodes are ASes and links are business relationships between ASes. That is, a link between a couple of ASes is a business agreement between the two organizations that enables these networks to exchange traffic. Recall that business relationships can be very roughly split into two classes: \textit{provider-customer} and \textit{peering}. Providers announce all the destinations to their customers, and thus forward all the traffic that their customers forward to them. Peers mutually announce a limited set of destinations, typically just their own destinations and their customer networks. Providers often charge their customers using the 95th percentile measurement schema \cite{NORTONPEERING}, i.e.\ the cost of the service depends on the amount of traffic exchanged. Peering is usually free of charge (unless maintenance costs are considered). The setup of public peering is greatly simplified by Internet eXchange Points (IXPs), facilities where each participant AS can create a single peering connection to any other participants that accept to peer (\textit{open peering} policy), or to a specific subset of those (\textit{selective/restrictive peering} policy). In what follows we explain the main observations from Section \ref{sec:analysis} in view of these Internet-specific realities.

\subsection{Growth of the $k_{MAX}$-dense index}\label{sec:kmax-growth}

Even though the size of the $H_{k_{MAX}}$ core fluctuates over time, Table~\ref{tab:avgkmax}, the $k_{MAX}$-dense index steadily grows, Figure \ref{fig:internetgrowth}. In this section, we provide some evidence that this growth is correlated with (if not due to) the proliferation of IXPs.

The first piece of evidence is the structural change of the Internet topology in the last decade due to different growth rates of peering and provider-customer relationships. The original economic driver behind peering is to reduce costs that customers pay to their providers. However, as the price of transit has been steadily decreasing, this driver becomes less prominent, unless large volumes of traffic are exchanged. As shown in \cite{PEERINGECONOMIES}, peering grows fast, contributing much more to middle AS tiers, compared to the tier-1 ASes. Large Content Providers (CP) and Content Delivery Networks (CDN), which generate a significant percentage of the total Internet traffic \cite{PEERINGECONOMIES,TRAFFIC} are primary drivers behind this process because:
\begin{enumerate}
\item a shorter path between these networks and subscribers provides better performances;
\item although the traffic exchanged is highly asymmetric, for most ISPs the connection to the content providers is vital.
\end{enumerate}
Furthermore, \cite{TRAFFIC} asserts that a dominant percentage of AS-level traffic flows directly between large CPs, CDNs, data centers, and consumer networks, and that $150$ ASes originate more than $50\%$ of the Internet inter-AS traffic. Reference \cite{LARGEIXP} also supports our claim by describing the ground truth behind a large European IXP: the authors shows that the amount of peering links established in this facility is unexpectedly huge---about $50,000$ peering links among approximately $350$ AS members. Yet another piece of evidence can be found in \cite{OURDENSE}: the percentage of ASes that are members of at least one IXP within a given $k$-dense set is a rapidly growing function of $k$. In addition, all the $k_{MAX}$-dense ASes are members of at least one IXP.

To further support our claim here we perform analysis similar to \cite{OURDENSE}. We collect the May 2012 information about the $60$ $k_{MAX}$-dense ASes in our 2012 snapshot from PeeringDB \cite{PDB}, a project maintaining a database aimed at facilitating the exchange of information related to peering: ``\textit{what networks are peering, where they are peering, and if they are likely to peer with you}''. We find that $58$ out of the $60$ ASes have a peering record in the database, and that \emph{all} these $58$ ASes are members of the Deutscher Commercial Internet Exchange (DE-CIX) \cite{DECIX}, one of the largest Internet Exchange Points worldwide, located in Frankfurt, Germany, with the current membership count of more than $450$ ASes. We investigate the profile of the remaining $2$ ASes, and find that they both are also DE-CIX members: one is a telecommunication company with an unknown peering policy, the other is a hosting service provider with an open peering policy.

\subsection{The $2$- and $3$-dense-sets}\label{sec:23-dense}

\begin{figure*}[t]
\centerline{\subfloat[]{\includegraphics[height=2cm]{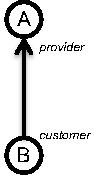}%
\label{subfig:pattern1}}
\hfil
\subfloat[]{\includegraphics[height=2cm]{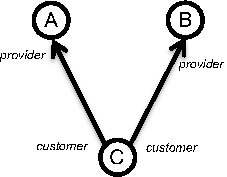}%
\label{subfig:pattern2}}
\hfil
\subfloat[]{\includegraphics[height=2cm]{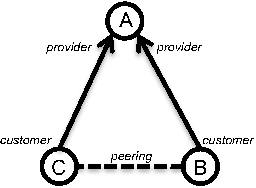}%
\label{subfig:pattern3}}
\hfil
\subfloat[]{\includegraphics[height=2cm]{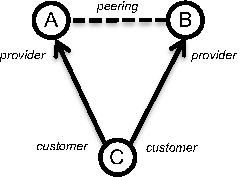}%
\label{subfig:pattern4}}
\hfil
\subfloat[]{\includegraphics[height=2cm]{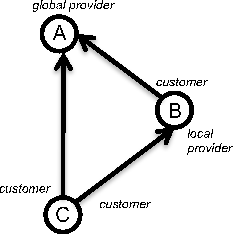}%
\label{subfig:pattern5}}}
\caption{Connectivity patterns of ASes with $k$-dense-indices equal to $2$ and $3$.}
\label{fig:2-3-conn}
\end{figure*}

Nodes with the \textit{k}-dense-index equal to $2$ or $3$ are a vast majority of all ASes in the Internet. These peripheral ASes contribute most to the overall network growth: they have small degrees, but a large percentage of all connections in the Internet are attached to these ASes \cite{ASEVOLUTION,TWELVE}. All ASes of degree $1$ and all ASes with zero clustering belong to the \textit{2}-dense-set. These customer ASes connect to the Internet, but their business is not related to Internet operations: they set up their connections to obtain Internet connectivity, thus all they need is a transit provider. Sometimes, for backup, higher availability, or other purposes, they set up agreements with multiple provider, i.e.\ they purchase transit from more than one provider (multihoming). If the two providers of a multihomed AS happen to be connected, then a triangle is formed, so that the multihomed AS belongs to the \textit{3}-dense-set, Figure~\ref{fig:2-3-conn}.

\subsection{The $k_{MAX}$-dense-set vs.\ high-rank ASes}\label{sec:kmax-dense}

\begin{table*}
\tiny
\centering
\caption{AS properties from PeeringDB. The \textit{$k_{MAX}$-dense} column refers to the $60$ $k_{MAX}$-dense ASes in the 2012 snapshot, while the \textit{degree-rank} and \textit{AS-rank} columns refer to the $60$ highest-degree and highest-AS-rank ASes~\cite{ASRANKFILE}. The data for the highest-address- and prefix-rank ASes are similar to the highest-AS-rank ASes.}
\label{tab:kmaxlabels}
\subfloat[][\label{subtab:bustype}Business type.]{
\begin{tabular}{l|c|c|c}
& $k_{MAX}$-dense & degree-rank & AS-rank \\
\hline
\hline
Cable/DSL/ISP		&	12  & 9 	& 5 \\
Content				&	4  	& 1 	& 0 \\
Educational/Research &	1 	& 3 	& 1 \\
Network Service Provider					&	42 	& 42 	& 43 \\
Non-Profit			&	1 	& 0 	& 0 \\
\hline
\end{tabular}
}
\qquad
\subfloat[][\label{subtab:geotype}Geographic type.]{
\begin{tabular}{l|c|c|c}
& $k_{MAX}$-dense & degree-rank & AS-rank \\
\hline
\hline
Asia Pacific	& 	0	& 0	 & 2  \\
Europe			& 	28 	& 20 & 8  \\
North America	&	0 	& 2	 & 5  \\
Regional		& 	17  & 8	 & 4  \\
Global			& 	15 	& 25 & 30 \\
\hline
\end{tabular}
}
\\
\subfloat[][\label{subtab:trvolume}Traffic volume.]{
\begin{tabular}{l|c|c|c}
& $k_{MAX}$-dense & degree-rank & AS-rank \\
\hline
\hline
100 Mbps - 10 Gbps	&	10 	& 6		& 0 \\
10 Gbps - 1 Tbps	&	41	& 36	& 29 \\
1Tbps+				&	2	& 7		& 12 \\
Unknown				& 	7	& 11	& 19 \\
\hline
\end{tabular}
}
\qquad
\subfloat[][\label{subtab:trratio}Traffic ratio.]{
\begin{tabular}{l|c|c|c}
& $k_{MAX}$-dense & degree-rank & AS-rank \\
\hline
\hline
Balanced			&	30	&	42	& 37 \\
Mostly Inbound		& 	7	& 	6	& 9  \\
Mostly Outbound		&	21	& 	7	& 3	 \\
Unknown				&	2	&	5	& 11 \\
\hline
\end{tabular}
}
\qquad
\subfloat[][\label{subtab:ppolicy}Peering policy.]{
\begin{tabular}{l|c|c|c}
& $k_{MAX}$-dense & degree-rank & AS-rank \\
\hline
\hline
Open		& 		32 	& 15	& 7		\\
Restrictive	& 		1 	& 11	& 15	\\
Selective	& 		26 	& 29	& 27	\\
Unknown		& 		1 	& 5		& 11 	\\
\hline
\end{tabular}
}
\end{table*}

The $k_{MAX}$-dense ASes form the densest-connected community by definition. The easiest way to establish such dense connectivity in practice is by connecting to a large IXP and declaring an open peering policy. Given that CPs and CDNs benefit from peering with any willing-to-peer ASes \cite{NORTONCP}, it is quite plausible that CPs and CDNs are main players behind the formation of this densely interconnected substructure. Surprisingly, the adoption of an open peering policy is an emerging phenomenon among Network Service Providers (NSPs) \cite{AMOGHPEERING}, i.e. tier-2 ASes provided with an own backbone network that purchase transit from an upstream provider and resell it to other ASes. Although these ASes usually adopt a selective peering policy, as they do not want to peer with potential customers, such peering connections help tier-2 ASes to provide a better end-user experience to their customers \cite{NORTONMOT}.

\begin{figure}
\includegraphics[width=0.5\textwidth]{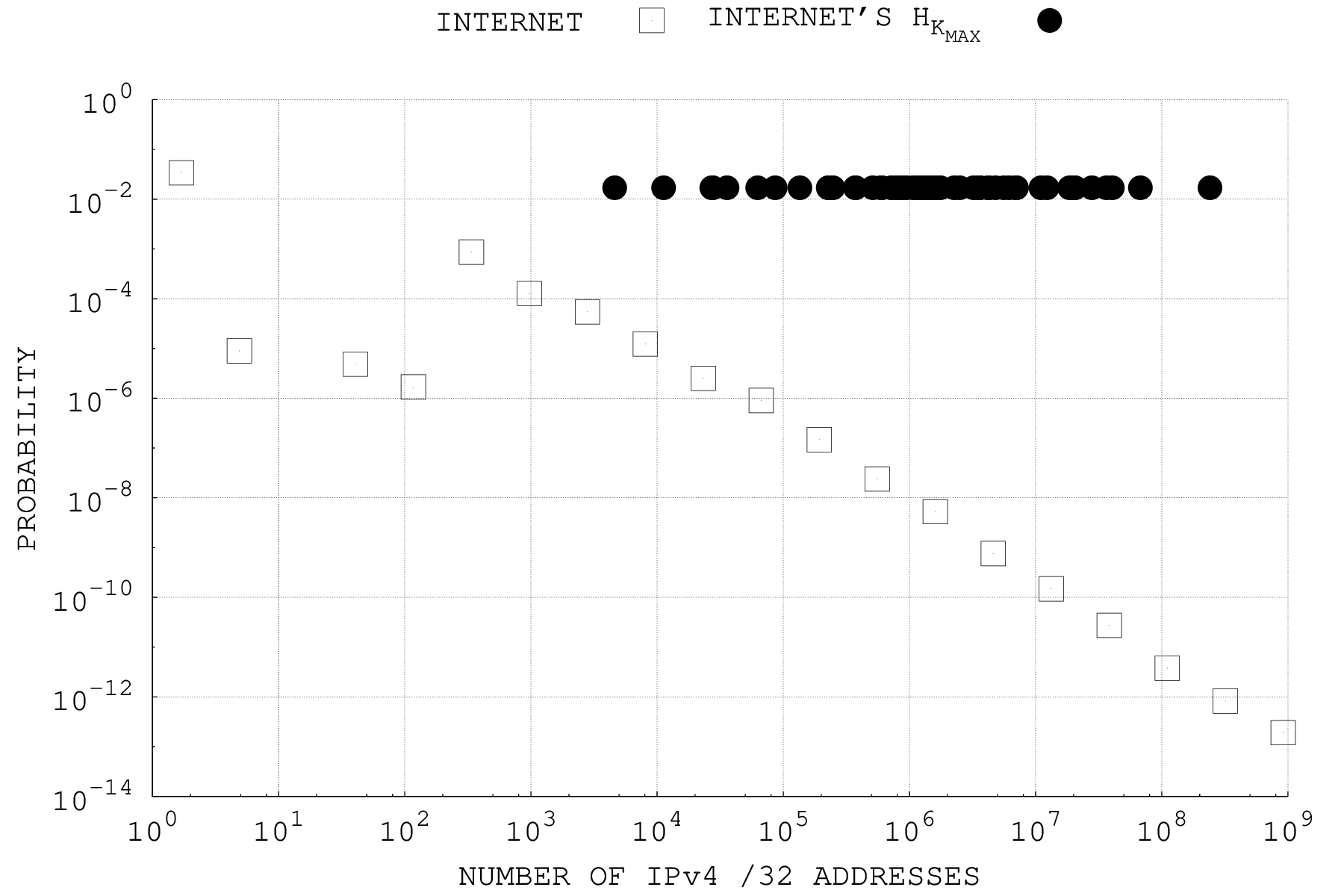}
\caption{Distributions of customer cone sizes for the $k_{MAX}$-dense ASes (black circles) and all ASes in the Internet (empty squares).}
\label{fig:ad}
\end{figure}

The PeeringDB data shown in Tables \ref{tab:kmaxlabels}\subref{subtab:bustype} and \ref{tab:kmaxlabels}\subref{subtab:trratio} confirm this statement. Indeed, large percentages of the $k_{MAX}$-dense ASes:
\begin{itemize}
\item can be considered as Content or Network Service Providers,
\item direct most of their traffic outbound, and
\item have an open peering policy.
\end{itemize}
%ASes with an open peering policy: 4 content, 18 NSPs, 8 access networks, 1 non-profit, 1 Educational/Research.
The percentage of ASes with selective peering policies (almost all NSPs) is also significant, but all these ASes are good candidates for selective peering as well, explaining their high $k$-density. Indeed, one commonly considered aspect in the peer selection process is the symmetry of the exchanged traffic \cite{NORTONFP}. We do not have access to traffic statistics, but we can use the number of IP addresses in the customer cone \cite{ASRANK} as a proxy. The customer cone of an AS is the set of ASes that can be reached from the AS following only provider-to-customer links. In other words, it is the set of destinations that can be reached \textit{for free} upon peering with the AS.
The distributions of the customer cone sizes (the numbers of /32 IP addresses in the customer cone, to be precise) shown in Figure \ref{fig:ad} indicate that the customer cone sizes of $k_{MAX}$-dense ASes are large, but their distribution is narrower that in the rest of the Internet, thus making $k_{MAX}$-dense ASes potential peer candidates even in the selective peering case.

We next juxtapose this $k_{MAX}$-dense-set against different sets of high-rank ASes, in particular high-degree ASes, shedding more light on the different Internet-specific meanings of the $k$-dense-index and node degree.

Tier-1 is defined as a set of ASes that do not need to pay any transit providers to reach any destination. To accomplish this, all the tier-1 ASes connect to each other forming a clique. This structure ensures that all these ASes have high, although not necessarily \emph{highest} \textit{k}-dense-indices. However, because they provide routes to all the destinations in the Internet without paying any upstream providers, they have \emph{largest} customer cones \cite{ASRANK,ASREL}.
Since the main business role of tier-1 ASes (or more generally, of any transit provider) is to sell transit, they have a lot of connections to enterprise customers, Figures \ref{subfig:2connections} and \ref{subfig:3connections}, but the open peering policy would not increase their revenue. Therefore, the tier-1 ASes are more likely to adopt restrictive peering policies. These considerations suggest that these ASes, contrary to a common belief, do not largely belong to the \emph{densest} community, i.e.\ to $H_{k_{MAX}}$.
In particular, the ASes with the largest customer cones are not $k_{MAX}$-dense, Figure \ref{fig:ad}.

In addition, we compute the overlaps between the set of the $60$ ASes in the $H_{k_{MAX}}$ core of the 2012 snapshot, and the top-$60$ ASes ranked by their customer cone size. The results in Table \ref{tab:asrank} show that these overlaps are not substantial. In the same table we also show the overlap between the $k_{MAX}$-max dense ASes and the top-$60$ highest-degree ASes. From Figure \ref{fig:dd2012} we know that a high degree does not necessarily mean a high \textit{k}-dense-index. In Table \ref{tab:asrank} we find that the overlap between these two AS sets is about $50\%$, and if we look back at Table \ref{tab:kmaxlabels}), we observe that these highest-degree or highest-rank ASes have quite different properties, compared to the $k_{MAX}$-dense ASes. Specifically, the former AS sets have
\begin{itemize}
\item a higher percentage of global ASes, Table \ref{tab:kmaxlabels}\subref{subtab:geotype},
\item a higher percentage of 10Gbps+ ASes, Table \ref{tab:kmaxlabels}\subref{subtab:trvolume},
\item a lower percentage of mostly outbound ASes, Table \ref{tab:kmaxlabels}\subref{subtab:trratio},
\item a higher percentage of ASes having a restrictive peering policy, Table \ref{tab:kmaxlabels}\subref{subtab:ppolicy}, and
\item a lower percentage of ASes having an open peering policy, Table \ref{tab:kmaxlabels}\subref{subtab:ppolicy}.
\end{itemize}
%All these statistics suggest that ASes in these ranks may have a different business profile, they are likely to be large transit providers. \\
Simply put, high-degree or high-rank ASes tend to be very large transit providers, while $k_{MAX}$-dense ASes tend to be either content providers or tier-2s.

\begin{table}
\centering
\caption{The $60$ $k_{MAX}$-dense ASes in the 2012 snapshot vs.\ top-$60$ high-rank ASes. The table shows the overlaps between the set of ASes in $H_{k_{MAX}}$ and: \textit{address-rank}, \textit{prefix-rank}, and \textit{AS-rank}: the sets of $60$ ASes with the largest numbers of IPv4 /32 addresses, IPv4 routing prefixes, and ASes in their customer cones; and \textit{degree-rank}: the set of $60$ highest-degree ASes. All the data are from~\cite{ASRANKFILE}.}
\label{tab:asrank}\begin{tabular}{c|c|c|c|c}
k-max dense & address-rank & prefix-rank & AS-rank & degree-rank \\
\hline
\hline
60    	&	4	&   6	&	7 & 31\\
\hline
\end{tabular}
\end{table}

\subsection{$1K$-randomness of the $H_{k_{MAX}}$ core}\label{sec:1k}

If an AS has an \textit{open peering policy}, it simply sets up a certain number of connections (equal to its degree), without \textit{choosing} its neighbors in any way, e.g.\ based on their degrees or any other properties. On the contrary, an AS with a \textit{selective/restrictive peering policy} always \textit{chooses} its peers and one of the topological property emerging from this selection process is the degree of the peering candidate, correlated with its customer cone size.
In the latter case, we thus cannot expect that the degree distribution alone is sufficient to fully describe the graph. Some non-trivial degree correlations must be present in it, and indeed the Internet AS-level graph as a whole was found to be not $1K$- but $2K$-random in~\cite{DKGRAPHS}.
The large percentage of open-peering $k_{MAX}$-dense ASes, and the similarity between their customer cone sizes are thus the main factors explaining the $1K$-randomness of the $H_{k_{MAX}}$ subgraph.

Let us assume for a moment that a large fraction of $k_{MAX}$-dense links is present at a single large IXP. We do not have access to the peering matrix data of any large IXPs, but the data reported in \cite{LARGEIXP} suggest that this assumption may very well be correct, with DE-CIX being one of possible large IXP candidates. If we also assume that the peering policies declared in PeeringDB reflect the peering policies that each AS follows at each IXP where it has presence, then our explanation of $H_{k_{MAX}}$'s $1K$-randomness is supported by the data in Figure~\ref{fig:ad} and Table~\ref{tab:kmaxlabels}\subref{subtab:ppolicy}.

\section{Conclusions}
\label{sec:conclusions}

In summary, as the Internet grows over time, the maximum $k$-dense index grows as well. That is, the densest and innermost Internet core $H_{k_{MAX}}$ becomes increasingly denser. Yet this form of densification can only partially be attributed to the growing average degree. Even though the node degrees and $k$-dense-indices are correlated, the relative fluctuations are strong, and the two statistics reflect two different Internet-specific properties of AS nodes. High-degree and high-$k$-dense-index ASes tend to be transit and content providers, respectively. The latter form the densest community in the Internet, which only loosely overlaps with the high-rank ASes forming the tier-1 core. The structure of the  $H_{k_{MAX}}$ core is relatively simple. Statistically, it is almost fully determined by its degree distribution, a property that can be explained by open peering policies that many content providers and NSPs tend to follow at IXPs. Most importantly, after proper normalization, all the considered $k$-dense properties of the Internet appear time-invariant. In particular, a vast majority of all AS links in the Internet are attached either to ASs with the $k$-dense-index equal to $2$ or $3$, or to the $k_{MAX}$-dense ASes.

Speaking more generally, the Internet's $k$-dense properties, derived from a recursive variant of edge multiplicity measuring the frequency and density of triangle overlaps, appear to be a statistically significant and time-invariant structural properties that cannot be fully captured by either degree distribution ($1K$-distribution) or degree correlations ($2K$-distribution). The $3K$-distribution, the distribution of subgraphs of size $3$, may very well capture the Internet's $k$-dense properties. Even though this conjecture is quite plausible, it is not guaranteed to be correct because reproducing the frequency of degree-labeled triangles does not automatically imply that the whole $k$-dense decomposition hierarchy of nested subgraphs $H_k$ is correctly reproduced as well. Therefore it is interesting to investigate what existing or new Internet topology models and generators are capable of explaining or at least reproducing the $k$-dense properties of the Internet. In that respect the recent results in~\cite{POLCLUST} are interesting as they indicate that some basic $k$-dense properties of the Internet and other networks are fairly similar to those in random graphs that have the same degree and {\em clustering\/} distributions. A more detailed analysis is needed to see how accurately such graphs reproducing clustering capture the full spectrum of the $k$-dense properties considered here. Yet it would not be too surprising if such an analysis does find that clustering indeed defines fairly accurately all the $k$-dense properties. Such results would not be surprising because clustering and $k$-dense properties are closely related, while the specifics of clustering are consequences of similarity forces driving network evolution, and taking these forces into account predicts remarkably well not only a long list of structural properties of different networks including the Internet, but also many details of the dynamics of their growth~\cite{POPSIM}.

\ifCLASSOPTIONcompsoc
  \section*{Acknowledgments}
\else
  \section*{Acknowledgment}
\fi

The authors would like to thank Amogh Dhamdhere, Alessandro Improta, Matthew Luckie and Luca Sani for sharing their data, and Hyunju Kim and Zoltan Toroczkai for sharing their code.

\ifCLASSOPTIONcaptionsoff
  \newpage
\fi




\bibliographystyle{IEEEtran}
\bibliography{bibliography}

\begin{IEEEbiography}[{\includegraphics[width=1in,height=1.25in,clip,keepaspectratio]{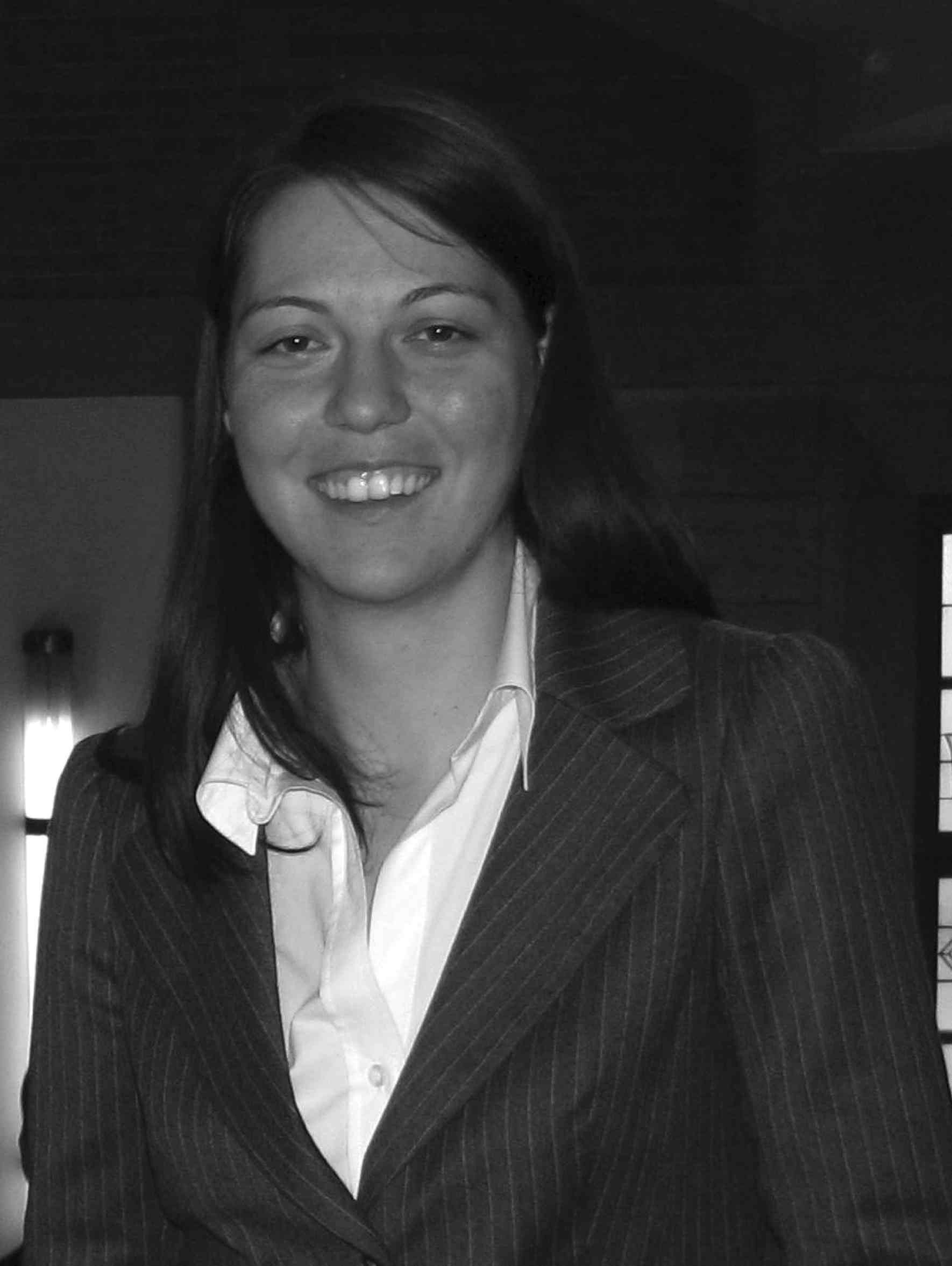}}]%
{Chiara Orsini}  received the Master’s and Ph.D. degrees in computer engineering from the University of Pisa, Pisa, Italy, in 2009 and in 2013, respectively. She is a Post-Doctoral Researcher with the Co- operative Association for Internet Data Analysis (CAIDA), La Jolla, CA, USA.  Her research interests focus primarily on studying the structural characteristics of the Internet topology at the AS-level of abstraction.
\end{IEEEbiography}

\vspace{-2cm}

\begin{IEEEbiography}[{\includegraphics[width=1in,height=1.25in,clip,keepaspectratio]{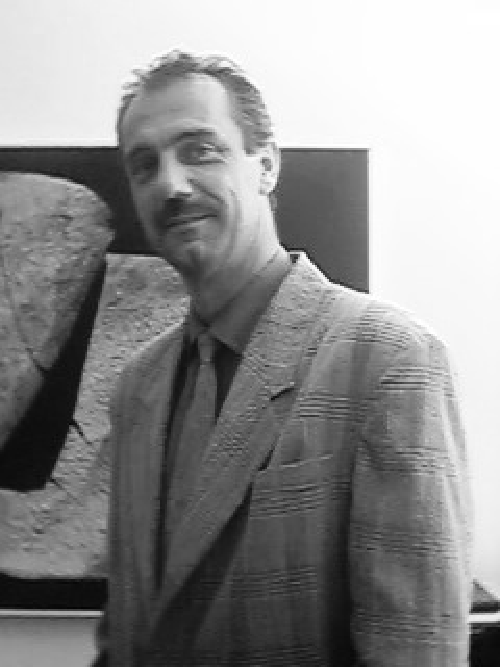}}]%
{Enrico Gregori} received the Laurea degree in electronic engineering from the University of Pisa, Pisa, Italy, in 1980. In 1981, he joined the Italian National Research Council (CNR), Pisa, Italy, where he is currently a CNR Research Director.  His current research interests include ad hoc networks, sensor networks, wireless LANs, quality of service in packet-switching networks, and evolution of TCP/IP protocols.  He has contributed to several national and international projects on computer networking. He has authored more than 100 papers in the area of computer networks, and he has published in international journals and conference proceedings.
\end{IEEEbiography}

\newpage

\begin{IEEEbiography}[{\includegraphics[width=1in,height=1.25in,clip,keepaspectratio]{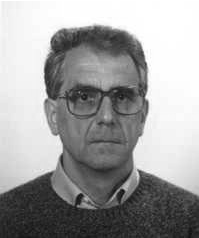}}]%
{Luciano Lenzini} received the Laurea degree in physics from the University of Pisa, Pisa, Italy, in 1969. He is a Full Professor in computer networking with the Faculty of Engineering, University of Pisa. His current research interests include the design and performance evaluation of MAC protocols for wireless networks and the quality of service provision in integrated and differentiated services networks. He is the author and coauthor of a high number of papers pub- lished in journals and conference proceedings. He is the author of numerous industrial patents. He has directed several national an international projects in the area of computer networking.
\end{IEEEbiography}

\vspace{-12cm}

\begin{IEEEbiography}[{\includegraphics[width=1in,height=1.25in,clip,keepaspectratio]{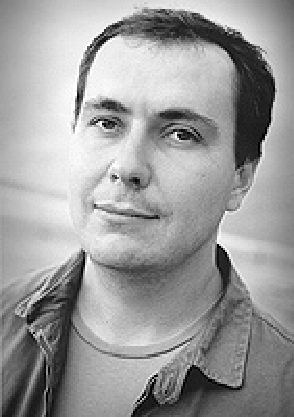}}]%
{Dmitri Krioukov} received the Diploma in Physics from the St. Petersburg State University, St. Petersburg, Russia, in 1993, and the Ph.D. degree in physics from Old Dominion University, Norfolk, VA, in 1998. He started working in the networking industry with Dimension Enterprises, Herndon, VA, USA. Upon its acquisition in 2002 by Nortel Networks, he accepted a Research Scientist position with Nortel in Herndon, Virginia. In 2004, he moved back to academia, and since then he has been a Senior Research Scientist with the Cooperative Association for Internet Data Analysis (CAIDA), University of California, San Diego (UCSD), La Jolla, CA, USA.
\end{IEEEbiography}

\end{document}